\documentclass[a4paper, cleveref, autoref, thm-restate]{lipics-v2021}

\usepackage{stmaryrd}
\usepackage{nicefrac}
\usepackage{amsthm}
\usepackage{complexity}
\usepackage{tabularray}
\usepackage{tikz}
\usetikzlibrary{calc,decorations.pathmorphing}
\usepackage{xparse}
\usepackage{bm}
\usepackage{multicol}

\newif\ifarxiv
\arxivtrue

\ifarxiv
\title{Validity of contextual formulas (extended version)}
\else
\title{Validity of contextual formulas}
\fi

\author{Javier Esparza}{Technical University of Munich, Germany}{esparza@in.tum.de}{https://orcid.org/0000-0001-9862-4919}{}\author{Rubén Rubio}{Universidad Complutense de Madrid, Spain}{rubenrub@ucm.es}{https://orcid.org/0000-0003-2983-3404}{}

\authorrunning{J. Esparza and R. Rubio} 

\Copyright{Javier Esparza and Rubén Rubio} 



\category{} 

\ifarxiv
\hideLIPIcs
\relatedversion{Extended version of the CONCUR 2024 paper~\cite{lipicsVersion}.} \else
\relatedversiondetails[linktext={arXiv:XXXX.XXXXX}, cite=fullversion]{Extended version}{https://doi.org/10.48550/arXiv.XXXX.XXXXX}
\fi

\funding{This work was partially supported by the Agencia Estatal de Investigación (AEI) under project PID2019-108528RB-C22.}

\nolinenumbers

\EventEditors{Rupak Majumdar and Alexandra Silva}
\EventNoEds{2}
\EventLongTitle{35th International Conference on Concurrency Theory (CONCUR 2024)}
\EventShortTitle{CONCUR 2024}
\EventAcronym{CONCUR}
\EventYear{2024}
\EventDate{September 9--13, 2024}
\EventLocation{Calgary, Canada}
\EventLogo{}
\SeriesVolume{311}
\ArticleNo{11}

\usepackage{todonotes}

\newcommand\cninst{\kappa} \newcommand{\decon}[1]{#1^d}
\newcommand\devarf[1]{#1^*}
\newcommand\devarp[1]{#1^+}
\newcommand{\consub}[1]{\textit{CSub}({#1})}
\newcommand\ctxarg{\psi}
\newcommand\otherarg{\vartheta}
\NewDocumentCommand\ctxap{G{c}G{\ctxarg}}{\ensuremath{p_{\Context{#1}{#2}}}}

\newcommand\true{\mathrm{true}}
\newcommand\false{\mathrm{false}}
\newcommand\X{\mathbf{X}\,}
\newcommand\U{\,\mathbf{U}\,}
\renewcommand\W{\,\mathbf{W}\,}
\renewcommand\G{\mathbf{G}\,}
\newcommand\F{\mathbf{F}\,}
\newcommand\GF{\mathbf{GF}\,}
\newcommand\FG{\mathbf{FG}\,}
\renewcommand\A{\mathbf{A}\,}
\newcommand\AG{\mathbf{AG}\,}
\renewcommand\AP{\mathit{AP}} \newcommand\N{\mathbb{N}}
\newcommand\mcEX{\langle \cdot \rangle\,}
\newcommand\mcAX{[ \cdot ]\,}
\newcommand*\ctltomc[1]{\ensuremath{#1^\mu}}
\newcommand*\FV[1]{\ensuremath{\mathrm{FV}(#1)}}

\newcommand\spotv{\textsc{Spot} 2.11}
\newcommand\mint{\hat\eta}
\newcommand{\msigma}{\hat\sigma}
\newcommand\coloneq{:=}
\newcommand\Coloneq{:\coloneq}
\newcommand\valuation{\beta}
\newcommand\feval[2][\beta]{\llbracket #2 \rrbracket_{#1}}
\newcommand\mueval[2][\eta]{\llbracket #2 \rrbracket_{#1}}
\newcommand\boolx[1][]{\ensuremath{\mathbb{F}_{#1}}}
\newcommand\mucx[1][]{\ensuremath{\mathbb{F}}} \newcommand\boolc[1][]{\ensuremath{\mathbb{C}_{#1}}}
\newcommand\mucc[1][]{\ensuremath{\mathbb{C}}} 
\newcommand\hole{[\ ]}
\newcommand\sbapp{\bar\sigma}
\newcommand*\valbound[1]{\ensuremath{\subseteq}}
\newcommand{\context}{c}
\newcommand{\contextVars}{\ensuremath{C}}
\NewDocumentCommand\Context{mg}{{#1} \IfNoValueTF{#2}{}{[{#2}]}
}
\newcommand\contextSet{\ensuremath{\mathcal{C}}}

\ifarxiv
\newcommand{\appendixref}[2]{\cref{#1}}
\newcommand{\fullversionref}{\cref{sec:proofs}}
\else
\newcommand{\appendixref}[2]{Appendix #2 of \cite{fullversion}}
\newcommand{\fullversionref}{\cite{fullversion}}
\fi

\newenvironment{sketch}{\begingroup\renewcommand*{\proofname}{Proof sketch {\normalfont(full proof in \fullversionref)}}\begin{proof}}{\end{proof}\endgroup}

\begin{document}

\maketitle

\begin{abstract}
Many well-known logical identities are naturally written as equivalences between \emph{contextual formulas}. A simple example is the Boole-Shannon expansion $\Context{c}{p} \equiv (p \wedge \Context{c}{\true} ) \vee (\neg\, p \wedge \Context{c}{\false} )$, where $\Context{c}$ denotes an arbitrary formula with possibly multiple occurrences of a ``hole'', called a \emph{context}, and $\Context{c}{\varphi}$ denotes the result of ``filling'' all holes of $\Context{c}$ with the formula $\varphi$. Another example is the unfolding rule $\mu X.  \Context{c}{X} \equiv  \Context{c}{\mu X.  \Context{c}{X}}$ of the modal $\mu$-calculus. 

We consider the modal $\mu$-calculus as overarching temporal logic and, as usual, reduce the problem whether $\varphi_1 \equiv \varphi_2$ holds for contextual formulas $\varphi_1, \varphi_2$  to the problem whether$\varphi_1 \leftrightarrow \varphi_2$ is valid . We show that the problem whether a contextual formula of the $\mu$-calculus is valid for all contexts can be reduced to validity of ordinary formulas.  Our first result constructs a \emph{canonical context} such that a formula is valid for all contexts if{}f it is valid for this particular one. However, the ordinary formula is exponential in the nesting-depth of the context variables. In a second result we solve this problem, thus proving that validity of contextual formulas is \EXP-complete, as for ordinary equivalences. We also prove that both results hold for CTL and LTL as well. We conclude the paper with some experimental results. In particular, we use our implementation to automatically prove the correctness of a set of  six contextual equivalences of LTL recently introduced by Esparza et al.\ for the normalization of LTL formulas. While Esparza et al. need several pages of manual proof, our tool only needs milliseconds to do the job and to compute counterexamples for incorrect variants of the equivalences.
\end{abstract}

\section{Introduction}

Some well-known identities useful for reasoning in different logics can only be easily formulated as \emph{contextual identities}. One example is the Boole-Shannon expansion of propositional logic, which constitutes the foundation of Binary Decision Diagrams and many SAT-solving procedures \cite{BiereHMW21}. It can be formulated as
\begin{equation}\label{eq:shannon}\Context{c}{p} \equiv (p \wedge  \Context{c}{\true}) \vee (\neg\, p \wedge  \Context{c}{\false})
\end{equation}
where, intuitively, $\Context{c}$ denotes a Boolean formula with ``holes'', called a \emph{context}, and $\Context{c}{\varphi}$ denotes the result of ``filling'' every hole of the context $\Context{c}$ with the formula $\varphi$. For example, if  $\Context{c} := (\hole  \wedge p) \vee (q \to  \hole)$ and $\varphi:= p$, then $\Context{c}{p} = (p  \wedge p) \vee (q \to p)$. More precisely, $\Context{c}$ is a \emph{context variable} ranging over contexts,  and the equivalence sign $\equiv$ denotes that for all possible assignments of contexts to $\Context{c}$ the ordinary formulas obtained on both sides of $\equiv$ are equivalent.

For linear-time temporal logic in negation normal form, a useful identity similar to the Boole-Shannon expansion is
\begin{equation}
\label{eq:gf}
  \Context{c}{\GF p} \equiv (\GF p \wedge  \Context{c}{\true}) \vee  \Context{c}{\false}
\end{equation}
\noindent where $\Context{c}$ now ranges over formulas of LTL with holes. For example, the identity shows that $q \U (\GF p \wedge r)$ is equivalent to $\GF p \wedge q \U r \vee q \U \false$, and so after simplifying equivalent to $\GF p \wedge q \U r$.\footnote{The restriction to formulas in negation normal form is necessary. For example, taking $\Context{c} := \neg (p \wedge \hole)$ does not yield a valid equivalence.} As a third example, the unfolding rule of the $\mu$-calculus (the fundamental rule in Kozen's axiomatization of the logic \cite{handbookMucalc}
\begin{equation*}
\mu X.  \Context{c}{X} \equiv  \Context{c}{\mu X. \Context{c}{X}}
\end{equation*}
\noindent whose formulation requires nested contexts. Further examples of contextual LTL identities are found in~\cite{jacm}, where, together with Salomon Sickert, we propose a rewrite system to transform arbitrary LTL formulas into formulas of the syntactic fragment $\Delta_2$, with at most single-exponential blowup.\footnote{$\Delta_2$ contains the formulas in negation normal form such that every path of the syntax tree exhibits at most one alternation of the strong and weak until operators $\U$ and $\W$. They have different uses, and in particular they are easier to translate into deterministic $\omega$-automata \cite{jacm}.} The rewrite system consists of the six identities (oriented from left to right) shown in Figure \ref{fig:rewrite}.

Remarkably, to the best of our knowledge the \emph{automatic} verification of contextual equivalences like the ones above has not been studied yet. In particular, we do not know of any automatic verification procedure for any of the identities above. In~\cite{jacm} we had to prove manually that the left and right sides of each identity are equivalent for every context (Lemmas 5.7, 5.9, and 5.11 of \cite{jacm}), a tedious and laborious task; for example, the proof of the first identity alone takes about $\nicefrac{3}{4}$ of a page. This stands in sharp contrast to non-contextual equivalences, where an ordinary equivalence $\varphi_1 \equiv \varphi_2$ of LTL can be automatically verified by constructing a Büchi automaton for the formula $\neg\, (\varphi_1 \leftrightarrow \varphi_2)$ and checking its emptiness. So the question arises whether the equivalence problem for contextual formulas is decidable, and in particular whether the manual proofs of \cite{jacm} can be replaced by an automated procedure. In this paper we give an affirmative answer.

\begin{figure}[t]
\begin{equation*}
\begin{array}{rcl}
\Context{c}{\psi_1 \U \psi_2} \W \varphi &  \equiv & (\GF \psi_2 \wedge \Context{c}{\psi_1 \W \psi_2} \W \varphi) \vee \Context{c}{\psi_1 \U \psi_2} \U (\varphi \vee \G \Context{c}{\false})  \\[1ex]
\varphi \W \Context{c}{\psi_1 \U \psi_2} & \equiv & \varphi \U \Context{c}{\psi_1 \U \psi_2} \vee \G \varphi  \\[1ex]
\Context{c}{\GF \psi} & \equiv & (\GF \psi \wedge \Context{c}{\true}) \vee \Context{c}{\false} \\[1ex]
\Context{c}{\FG \psi} & \equiv & (\FG \psi \wedge \Context{c}{\true}) \vee \Context{c}{\false}  \\[1ex]
\GF \Context{c}{\psi_1 \W \psi_2} & \equiv & \GF \Context{c}{\psi_1 \U \psi_2} \vee  (\FG \psi_1 \wedge \GF \Context{c}{\true})  \\[1ex]
\FG \Context{c}{\psi_1 \U \psi_2} & \equiv & (\GF \psi_2 \wedge \FG \Context{c}{\psi_1 \W \psi_2}) \vee \FG \Context{c}{\false}  \\[1ex]
\end{array}
\end{equation*}
\caption{Rewrite system for the normalization of LTL formulas~\cite{jacm}.}
\label{fig:rewrite}
\end{figure}

Let $\sigma$ be a mapping assigning contexts to all context variables of a formula $\varphi$, and let $\sigma(\varphi)$ denote the ordinary formula obtained by instantiating $\varphi$ with $\sigma$. The equivalence, validity, and satisfiability problems for contextual formulas are:
\begin{enumerate}
\item \emph{Equivalence} of $\varphi_1$ and $\varphi_2$: does $\sigma(\varphi_1) \equiv \sigma(\varphi_2)$ hold for every $\sigma$?
\item \emph{Validity} of $\varphi$: is $\sigma(\varphi)$ valid (in the ordinary sense) for every $\sigma$?
\item \emph{Satisfiability} of $\varphi$: is $\sigma(\varphi)$ satisfiable (in the ordinary sense) for some $\sigma$?
\end{enumerate}
We choose the modal $\mu$-calculus as overarching logic, and prove that these problems can be reduced to their counterparts for ordinary $\mu$-calculus formulas. As corollaries, we also derive reductions for CTL and LTL. More precisely, we obtain the following two results.

\subparagraph{First result.} Given a contextual formula $\varphi$ with possibly multiple occurrences of a context variable $\Context{c}$,  there exists a \emph{canonical instantiation $\Context{\cninst_\varphi}$} of $c$, also called the \emph{canonical context}, such that $\varphi$ is valid/satisfiable  if{}f the ordinary formula $\Context{\cninst_\varphi}(\varphi)$ is valid/satisfiable. Further, $\Context{\cninst_\varphi}$ can be easily computed from $\varphi$ by means of a syntax-guided procedure.

To give a flavour of the idea behind the canonical instantiation  consider the distributive law $\varphi_1 \wedge (\varphi_2 \vee \varphi_3) \equiv (\varphi_1 \wedge \varphi_2) \vee (\varphi_1 \wedge \varphi_3)$ for ordinary Boolean formulas. It is well-known that such a law is correct if{}f it is correct for the special case in which $\varphi_1, \varphi_2, \varphi_3$ are distinct Boolean variables, say $p_1, p_2, p_3$. In other words, the law is correct if{}f the Boolean formula $p_1 \wedge (p_2 \vee p_3) \leftrightarrow (p_1 \wedge p_2) \vee (p_1 \wedge p_3)$ is valid. This result does not extend to contextual formulas. For example, consider the contextual equivalence (\ref{eq:gf}),  reformulated as the validity of the contextual formula
\begin{equation}
\label{eq:gfvalid}
 \varphi :=  \Context{c}{\GF p} \; \leftrightarrow \; \big( ( \GF p \wedge  \Context{c}{\true}) \vee  \Context{c}{\false} \big)
 \end{equation}
While $\varphi$ is valid, the ordinary formula $\decon{\varphi}:= p_1 \leftrightarrow \big( (\GF p \wedge p_2) \vee p_3 \big)$ obtained by replacing $\Context{c}{\GF q}$, $\Context{c}{\true}$, and $\Context{c}{\false}$ by atomic propositions $p_1, p_2, p_3$, respectively,  is not. (We call $\decon{\varphi}$ the \emph{decontextualization} of $\varphi$.) Loosely speaking, the replacement erases dependencies between $\Context{c}{\GF q}$, $\Context{c}{\true}$, and $\Context{c}{\false}$. For example, since contexts are formulas in negation normal form, $\Context{c}{\false} \models \Context{c}{\true}$ or $\Context{c}{\false} \models \Context{c}{\GF p}$ hold for every context $c$,  but we do not have $p_3 \models p_2$ and $p_2 \models p_1$. To remedy this, we choose a context $\Context{\cninst_\varphi}$ that informally states:
\begin{quote}
At every moment in time, $p_1$ holds if the hole is filled with a formula globally entailing $\GF  p$ and $p_2$ holds if it is filled with a formula globally entailing $\true$ and $p_3$ holds if it is filled with a formula globally entailing $\false$.
\end{quote}
The context is:
\begin{equation}
\label{cancontext}
\cninst_{\varphi} := \G \bigg( \big(\G(\hole \to \GF p) \to p_1 \big) \wedge \big( \G(\hole \to \true) \to p_2 \big)   \wedge \big( \G(\hole \to \false) \to p_3) \big) \bigg)$$
\end{equation}
\noindent Our result shows that $\Context{\cninst_\varphi}$ is a canonical context for $\varphi$. In other words, the contextual formula $\varphi$ of (\ref{eq:gfvalid}) is valid if{}f the ordinary formula 
\begin{equation*}
\label{firstreduction}
\cninst_{\varphi}(\varphi) := \ \Context{\cninst_\varphi}{\GF p} \; \leftrightarrow \; \big( ( \GF p \wedge  \Context{\cninst_\varphi}{\true}) \vee \Context{\cninst_\varphi}{\false} \big)
 \end{equation*}
\noindent obtained by setting $c:=\cninst_{\varphi}$ in (\ref{eq:gfvalid}), is valid. After substituting according to (\ref{cancontext}) and simplifying, we obtain 
\begin{equation}
\label{firstreductionsimp}
\cninst_{\varphi}(\varphi) \equiv \big(\G(p_1 \wedge p_2) \wedge \G\big(\GF p \vee p_3) \big) \leftrightarrow (\GF p \wedge \G( (\GF p \to p_1) \wedge p_2) \vee \G(p_1\wedge p_2 \wedge p_3)\big)
 \end{equation}
 \noindent So (\ref{eq:gfvalid}) is valid if{}f the formula on the right-hand-side of (\ref{firstreductionsimp}) is valid, which is proved by \spotv{} \cite{spot} in milliseconds.

\subparagraph{Second result.} Given a contextual formula $\varphi$, the ordinary formula $\cninst_{\varphi}(\varphi)$ has  $O(|\varphi|^d)$ length, where $d$ is the nesting depth of the context variables.  Since $d \in O(n)$, the blowup is exponential. Our second result provides a polynomial reduction. Let $\Context{c}{\psi_1}$, \ldots, $\Context{c}{\psi_n}$ be the context expressions appearing in $\varphi$. Instead of
finding a canonical instantiation, we focus on adding to the decontextualized formula $\decon{\varphi}$ information on the dependencies between $\Context{c}{\psi_1}$, \ldots, $\Context{c}{\psi_n}$. For every pair $\psi_i, \psi_j$, we add to  $\decon{\varphi}$ the premise
$\G\big( \G(\psi_i  \to \psi_j) \to (p_i \to p_j)\big)$. Intuitively, the premise ``transforms'' dependencies between $\psi_i$ and $\psi_j$ into dependencies between fresh atomic propositions $p_i$ and $p_j$. For example, we obtain that (\ref{eq:gfvalid}) is valid if{}f the ordinary LTL formula

\begin{align*}
\label{secondreduction}
\G \left( \bigwedge_{i=1}^3 \bigwedge_{j=1}^3  \big(\G(\psi_i  \to \psi_j) \to (p_i \to p_j)\big) \right)  \to  \big( p_1 \leftrightarrow ((\GF p \wedge p_2) \vee p_3)
\end{align*}
\noindent is valid or, after simplification, if{}f
\begin{equation}
\label{red}
\begin{array}{c}
 \G \left(
 \begin{array}{c}
 (\FG \neg p \to (p_1 \to p_3)) \wedge (\GF p \to (p_2 \to p_1)) \\
 \wedge  \\
  (p_3 \to p_1) \wedge (p_1 \to p_2)
  \end{array}
  \right)
  \to
  \big( p_1 \leftrightarrow ((\GF p \wedge p_2) \vee p_3) \big)
 \end{array}
\end{equation}
\noindent is valid. Again, \spotv{} proves that (\ref{red}) is valid within milliseconds. Since the premise has polynomial size in the size of the original contextual formula, we obtain a reduction from contextual validity to ordinary validity with polynomial blowup. Observe, however, that the ordinary formula  is not obtained by directly instantiating the context variable $\Context{c}$.

\subparagraph{Experiments.} We have implemented our reductions and connected them to validity and satisfiability checkers for propositional logic (PySAT~\cite{pysat} and MiniSat~\cite{minisat}), LTL (\spotv~\cite{spot}), and CTL (CTL-SAT~\cite{ctlsat}). We provide some experimental results. In particular, we can prove the correctness of all the LTL identities of \cite{jacm} within milliseconds.

\subparagraph{Structure of the paper.}  \Cref{sec:preliminaries} recalls the standard $\mu$-calculus and presents its contextual extension. \Cref{sec:boolean} studies the validity problem of contextual propositional formulas, as an appetizer for the main results on the $\mu$-calculus in \cref{sec:mucalc}. These are extended to CTL and LTL in \cref{sec:ctl,sec:ltl}. Experimental results are presented in \cref{sec:experiments}, and \cref{sec:conclusions} gives some conclusions.

\section{The contextual \textmu-calculus} \label{sec:preliminaries}
We briefly recall the syntax and semantics of the $\mu$-calculus \cite{handbookMucalc}, and then introduce the syntax and semantics of the contextual $\mu$-calculus.
\subparagraph{The modal \textmu-calculus.} The syntax of the modal $\mu$-calculus over a set $\AP$ of atomic propositions and a set $V$ of variables is
\begin{equation}
\label{eq:syntax}
	\varphi \Coloneq p \mid \neg\, p \mid  X \mid \varphi \wedge \varphi \mid \varphi \vee \varphi \mid \mcEX \varphi \mid  \mcAX \varphi \mid \mu X. \varphi \mid \nu X.\varphi
\end{equation}
\noindent where $p \in \AP$ and $X \in V$.  The semantics  is defined with respect to a Kripke structure and a valuation.  A \emph{Kripke structure} is a tuple $\mathcal{K} = (S, {\to}, I, \AP, \ell)$, where $S$ and $I$ are sets of states and initial states, ${\to} \subseteq S \times S$ is the transition relation (where we assume that every state has at least one successor), and $\ell \colon S \to \mathcal{P}(S)$ assigns to each state a set of atomic propositions. A \emph{valuation} is a mapping $\eta \colon V \to \mathcal{P}(S)$.
Given $\mathcal{K}$ and $\eta$, the semantics assigns to each formula $\varphi$ a set $\mueval{\varphi} \subseteq S$, the set of states satisfying $\varphi$A. A Kripke structure $\mathcal{K}$ satisfies $\varphi$  if $I \subseteq \mueval{\varphi}$. The mapping $\mueval{\cdots}$ is inductively defined by:
\begin{center}
\begin{minipage}[c]{5cm}
\begin{align*}
	\mueval{p} & =  \{ s \in S \mid p \in \ell(s) \}  \\
	\mueval{\neg p} & = S \setminus \mueval{p}  \\
	\mueval{X} & = \eta(X) \\
\mueval{\varphi_1 \wedge \varphi_2} & = \mueval{\varphi_1} \cap \mueval{\varphi_2} \\
	\mueval{\varphi_1 \vee \varphi_2} & = \mueval{\varphi_1} \cup \mueval{\varphi_2}
\end{align*}
\end{minipage}
\begin{minipage}[c]{5cm}
\begin{align*}
	\mueval{\mcEX \varphi} & = \{ s \in S \mid \exists s' .  s \to s' \wedge s' \in \mueval{\varphi} \} \\
	\mueval{\mcAX \varphi} & = \{ s \in S \mid \forall s' .  s \to s' \Rightarrow s' \in \mueval{\varphi} \} \\
	\mueval{\mu X. \varphi} & = \bigcap \{ U \subseteq S \mid \mueval[{\eta[X/U]}]{\varphi} \subseteq U \} \\
	\mueval{\nu X. \varphi} & = \bigcup \{ U \subseteq S \mid U \subseteq \mueval[{\eta[X/U]}]{\varphi} \}
\end{align*}
\end{minipage}
\end{center}
Let $\FV{\varphi} \subseteq V$ be the set of free variables, i.e.\ not bound by a fixpoint operator, in a formula $\varphi$. Observe that if $\varphi$ is a closed formula (that is, $\FV{\varphi} = \emptyset$), then $\mueval{\varphi}$ depends only on $\mathcal{K}$, not on $\eta$, so we just write $\mueval[]{\varphi}$. On the contrary, when dealing with multiple Kripke structures at the same time, we write $\mueval[\mathcal{K},\eta]{\varphi}$ or $\mueval[\mathcal{K}]{\varphi}$ to avoid ambiguity. We say that $\mathcal{K}$ satisfies $\varphi$, denoted $\mathcal{K} \models \varphi$,  if $I \subseteq \mueval{\varphi}$, that is, if every initial state satisfies $\varphi$. A closed formula $\varphi$ is valid (satisfiable) if $\mathcal{K} \models \varphi$ for every (some) Kripke structure $\mathcal{K}$.

It is well-known that every formula of the $\mu$-calculus is equivalent to a formula in which all occurrences of a variable are either bound or free, and every two distinct  fixpoint subformulas have different variables.

\subparagraph{The contextual modal \textmu-calculus.}
\emph{Contextual formulas} are expressions over a set $\AP$ of atomic propositions, a set $V$ of variables, and a set $\contextVars$ of \emph{context variables}. (The contextual formulas of the introduction only had one contextual variable, but in general they can have multiple and arbitrarily nested variables.) They are obtained by extending the syntax (\ref{eq:syntax}) with a new term:
\begin{equation*} \varphi \Coloneq  p \mid \neg p \mid X \mid \cdots  \mid \nu X. \varphi  \mid \Context{c}{\varphi}
\end{equation*}
\noindent where $c \in \contextVars$. For the semantics, we need to introduce contexts and their instantiations. A \emph{context} is an expression over the syntax that extends (\ref{eq:syntax}) with \emph{holes}:
\begin{equation*} \varphi \Coloneq  p \mid \neg p \mid X \mid \cdots  \mid \nu X. \varphi  \mid \hole
\end{equation*}
\noindent We let $\contextSet$ denote the set of all contexts. An  \emph{instantiation} of the set $C$ of context variables is a mapping $\sigma \colon \contextVars \to \mathcal{C}$. Given a contextual formula $\varphi$, we let $\sigma(\varphi)$ denote  the ordinary formula obtained as follows: in the syntax tree of $\varphi$, proceeding bottom-up, repeatedly replace each expression $\Context{c}{\psi}$ by the result of filling all holes of the context $\sigma(c)$ with $\psi$. Here is a formal inductive definition:

\begin{definition}[instantiation]
Let $\mucx$ and $\mathbb{C}$ be the sets of ordinary and contextual formulas of the $\mu$-calculus. 
An \emph{instantiation} is a function $\sigma : \contextVars  \to \contextSet$ binding each context variable to a context. We lift an instantiation $\sigma$ to a mapping $\sbapp_c : \mathbb{C} \to \mucx$ as follows:
\begin{enumerate}
	\item $\sbapp_c(p) = p$.
	\item $\sbapp_c(\Context{c}{\varphi}) = (\sigma(c))[\hole/\sbapp_c(\varphi)]$ (i.e., the result of substituting $\sbapp_c(\varphi)$ for $\hole$ in $\sigma(c)$).
	\item $\sbapp_c(\varphi_1 \wedge \varphi_2) = \sbapp_c(\varphi_1) \wedge \sbapp_c(\varphi_2)$.
	\item $\sbapp_c(\varphi_1 \vee \varphi_2) = \sbapp_c(\varphi_1) \vee \sbapp_c(\varphi_2)$.
	\item $\sbapp_c(\mcEX \varphi) = \mcEX \sbapp_c(\varphi)$.
	\item $\sbapp_c(\mcAX \varphi) = \mcAX \sbapp_c(\varphi)$.
	\item $\sbapp_c(\mu X. \varphi) = \mu X. \sbapp_c(\varphi)$.
	\item $\sbapp_c(\nu X. \varphi) = \nu X. \sbapp_c(\varphi)$.
\end{enumerate}
Abusing language, we overload $\sigma$ and write $\sigma(\varphi)$ for $\sbapp(\varphi)$.
\end{definition}

\begin{example}
Let $\varphi = p \U \Context{c_1}{\;  (p \vee \Context{c_1}{q}) \W \Context{c_2}{\neg p \vee q)} \;}$. Further, let $\sigma(c_1) := \G \hole$ and $\sigma(c_2) := (\hole \wedge q)$. 
We have $\sigma(\varphi) = p \U \G\big((p \vee \G q) \W ((\neg p \vee q) \wedge q)\big)$.
\end{example}

We can now extend the notions of validity and satisfaction from ordinary to contextual formulas.

\begin{definition}[validity and satisfiability] \label{def:satisfaction}
A closed contextual formula $\varphi$ is valid if $\mathcal{K} \models \sigma(\varphi)$ for every instantiation $\sigma$ and Kripke structure $\mathcal{K}$, and satisfiable if 
$\mathcal{K} \models \sigma(\varphi)$ for some instantiation $\sigma$ and Kripke structure $\mathcal{K}$.
\end{definition}

\section{Validity of contextual propositional formulas} \label{sec:boolean}

As an appetizer, we study the validity and satisfiability problems for the propositional fragment of the modal $\mu$-calculus, which allows us to introduce the main ideas in the simplest possible framework.
The syntax of contextual propositional formulas over sets $\AP$ and $\contextVars$ of propositional and contextual variables is
\begin{equation}
\label{syntaxpropfragment}
\varphi \Coloneq p \mid \neg\, p \mid \varphi \wedge \varphi \mid \varphi \vee \varphi \mid \Context{c}{\varphi} 
\end{equation}
where $p \in \AP$ and $\context \in \contextVars$. The semantics is induced by the semantics of the modal $\mu$-calculus, but we quickly recall it. Given a \emph{valuation} $\beta \colon \AP \to \{0,1\}$, the semantics of an ordinary formula $\varphi$ is the Boolean $\feval{\varphi} \in \{0, 1\}$, defined as usual, e.g.\ $\feval{\varphi_1 \wedge \varphi_2}=1$ if{}f $\feval{\varphi_1}=1$ and $\feval{\varphi_2}=1$. Given a valuation $\beta$ and an instantiation $\sigma \colon \contextVars \to \contextSet$ of the context variables, the semantics of a contextual formula is the boolean $\feval{\sigma(\varphi)}$, where $\sigma(\varphi)$ is the formula obtained by instantiating each context variable $c$ with the context $\sigma(c)$.

We will use  the substitution lemma of propositional logic. Let $\boolx$ and $\boolc$ be the set of all ordinary and contextual propositional formulas, respectively.

\begin{lemma}[substitution lemma] \label{lem:bool:subs}
For any $\varphi \in \boolx$, valuation $\valuation \colon \AP \to \{0,1\}$, and substitution $\sigma \colon \AP  \to \boolx$,
$ \feval{\sigma(\varphi)} = \feval[\valuation']{\varphi} $
where $\beta'$ is given by $\valuation'(p) := \feval{\sigma(p)}$ for every $p \in V$.
\end{lemma}

\noindent Moreover, since formulas with syntax  (\ref{syntaxpropfragment}) are in negation normal form, we have the following monotonicity result.

\begin{lemma}[monotonicity] \label{lem:bool:monotonic}
For any $\varphi, \psi, \psi' \in \boolx$, propositional variable $p$ that does not appear negated in $\varphi$, and valuation $\beta \colon \AP \to \{0,1\}$, if $\feval{\psi \to \psi'}=1$ then $\feval{\varphi[p/\psi] \to \varphi[p/\psi']}=1$.
\end{lemma}

\subsection{Canonical instantiations} \label{sec:bool:method1}
Let us now prove that a contextual propositional formula $\varphi \in \boolc$ is valid (satisfiable) if{}f it is valid (satisfiable) for the following \emph{canonical instantiation} $\cninst_\varphi$ of its context variables.

\begin{definition}[maximal context subformulas]
A \emph{context subformula} of $\varphi \in \boolc$ is a subformula of $\varphi$ of the form $\Context{c}{\psi}$ for some $\Context{c} \in \contextVars$ and $\psi \in \boolc$.
The set of context subformulas of $\varphi$ is denoted $\consub{\varphi}$.
A context subformula  is \emph{maximal} if it is not a proper subformula of any other context subformula. The \emph{decontextualization} of $\varphi$, denoted $\decon{\varphi}$, is the result
of replacing every maximal context subformula $\Context{c}{\psi}$ of $\varphi$ by a fresh propositional variable $p_{\Context{c}{\psi}}$.
\end{definition}

\begin{definition}[canonical instantiation of a contextual formula]
\label{def:canonical}
The \emph{canonical instantiation} of $\varphi \in \boolc$, also called the \emph{canonical context}, is the mapping $\cninst_\varphi \colon \contextVars \to \contextSet$ that assigns to every context variable $\Context{c} \in \contextVars$ the context
\[ \cninst_\varphi(\Context{c}) \coloneq \bigwedge_{\Context{c}{\psi} \in \consub{\varphi}} \left( \hole \rightarrow \decon{\psi} \right) \to p_{\Context{c}{\psi}} \]
\end{definition}

\begin{example}
\label{ex:shannoncanonical}
Let us illustrate Definition \ref{def:canonical} on an example. Boole-Shannon's expansion holds if{}f the contextual formula
\begin{equation*}
\varphi :=  \Context{c}{p} \leftrightarrow \big( (p \wedge  \Context{c}{\true}) \vee (\neg\, p \wedge  \Context{c}{\false}) \big)
 \end{equation*}
 \noindent is valid. We have $\consub{\varphi} = \{ \Context{c}{p}, \Context{c}{\true},\Context{c}{\false}\}$. All elements of $\consub{\varphi}$ are maximal. Since $\decon{p} = p$, $\decon{\true} = \true$, and $\decon{\false} = \false$, the canonical instantiation is given by
 \begin{align*}
 \cninst_\varphi(\Context{c}) & = \big(  (\hole \to p) \to p_{\Context{c}{p}} \big) \wedge   \big(  (\hole \to \true) \to p_{\Context{c}{\true}} \big) \wedge \big(  (\hole \to \false) \to p_{\Context{c}{\false}} \big)
 \end{align*}
\end{example}

We need an auxiliary lemma.

\begin{restatable}{lemma}{boolModelprime} \label{lem:bool:modelprime}
For any $\varphi \in \boolc$, instantiation $\sigma$, and valuation $\valuation$, there is a valuation $\valuation'$ that coincides with $\valuation$ in every variable occurring in $\sigma(\varphi)$ and satisfies $\feval[\valuation]{\sigma(\varphi)} = \feval[\valuation']{\cninst_\varphi(\varphi)}$.
\end{restatable}

\begin{sketch}
Given $\varphi$, $\sigma$, and $\valuation$, we define $\valuation'(\ctxap) = \feval{\sigma(\Context{c}{\ctxarg})}$ for every $\Context{c}{\ctxarg} \in \consub{\varphi}$ and $\valuation'(p) = \valuation(p)$ otherwise. We first prove $\feval{\sigma(\phi)} = \feval[\valuation']{\decon{\phi}}$ as a direct application of the substitution lemma with $\gamma_{\sigma}(\ctxap) = \sigma(\Context{c}{\ctxarg})$, which satisfies $\gamma_{\sigma}(\decon{\phi}) = \sigma(\phi)$. Then, we prove $\feval{\sigma(\phi)} = \feval[\valuation']{\cninst_\varphi(\phi)}$ by induction on $\phi$, using the previous statement and some calculations on the expression of $\cninst_\varphi(\Context{c}{\ctxarg})$ using the monotonicity of contexts by \cref{lem:bool:monotonic}.
\end{sketch}

\begin{proposition}[fundamental property of the canonical instantiation] \label{thm:bool:main}
A contextual formula $\varphi \in \boolc$ is valid (resp.\ satisfiable) if{}f the ordinary formula $\cninst_\varphi(\varphi) \in \boolx$ is valid (resp. satisfiable).
\end{proposition}

\begin{proof}
For validity, if $\varphi$ is valid, then $\sigma(\varphi)$ is valid for every instantiation $\sigma$, and so in particular $\cninst_\varphi(\varphi)$ is valid.
For the other direction, assume $\cninst_\varphi(\varphi)$ is valid. We prove that  $\sigma(\varphi)$ is also valid for any instantiation $\sigma$. Let $\valuation$ be a valuation. By \cref{lem:bool:modelprime} there is another valuation $\valuation'$ such that $\feval{\sigma(\varphi)} = \feval[\valuation']{\cninst(\varphi)}$. Moreover, we have $\feval[\valuation']{\cninst(\varphi)} = 1$ because $\cninst(\varphi)$ is valid. So $\sigma(\varphi)$ is valed,  because $\beta$ is arbitrary.

Satisfiability is handled by a dual proof. If $\cninst_\varphi(\varphi)$ is satisfiable, then so is $\varphi$ by definition. If $\varphi$ is satisfiable, then there is an instantiation $\sigma$ such that $\feval{\sigma(\varphi)} = 1$. \Cref{lem:bool:modelprime} give us a valuation $\valuation'$ such that $\feval[\valuation']{\cninst_\varphi(\varphi)} = \feval{\sigma(\varphi)} = 1$.
\end{proof}

\begin{example}
\label{ex:shannon2}
Let $\varphi$ and $\cninst_\varphi(\Context{c})$ be as in  Example \ref{ex:shannoncanonical}. 
 By definition, we have $\ \cninst_\varphi(\varphi)  :=  \cninst_\varphi(\Context{c}{p})  \leftrightarrow  \big( (p \wedge  \cninst_\varphi(\Context{c}{\true})) \vee (\neg\, p \wedge  \cninst_\varphi(\Context{c}{\false})) \big)$.
Simplification yields
 $$\begin{array}{rcc}
& &   p_{\Context{c}{p}}    \wedge     p_{\Context{c}{\true}}   \wedge     (\neg p \to p_{\Context{c}{\false}})    \\
\cninst_\varphi(\varphi)  & \equiv & \leftrightarrow \\
& &  (p \wedge (p \to p_{\Context{c}{p}}) \wedge p_{\Context{c}{\true}}) \vee
 (\neg p \wedge p_{\Context{c}{p}} \wedge p_{\Context{c}{\true}} \wedge p_{\Context{c}{\false}})
\end{array}$$
This formula is not valid, and so by Proposition \ref{thm:bool:main} Boole-Shannon's expansion is valid.
\end{example}

The following example shows that the ordinary formula $\cninst_\varphi(\varphi)$ may be exponentially larger than the contextual formula $\varphi$ when $\varphi$ contains nested contexts. 

\begin{example} \label{example:nestedblowup}
Consider the contextual formula $\varphi := c^n[q]$, where $c^0[\psi] := \psi$ and $c^n[\psi] := \Context{c}{\Context{c^{n-1}}{\psi}}$ for every formula $\psi$. The size of $\varphi$ is $n + 4$. The canonical context is $\cninst_\varphi(c) = \bigwedge_{l=1}^n (\hole \to p_{\Context{c^{l-1}}{q}}) \to p_{\Context{c^l}{q}}$ with $p_{\Context{c^0}{q}} = q$. Instantiating the ``holes'' of $\cninst_\varphi(c)$ with a formula $\psi$ of size $k$ yields the formula $\cninst_\varphi(c)[ \hole/\psi]$ of size $n (7 + k) - 1 \geq n k$. Since $\cninst_\varphi(\varphi) = \cninst_\varphi(c^n[q]) = \cninst(c)[\hole/\cninst(c^{n-1}[q])]$ by definition, the size of  $\cninst_\varphi(\varphi)$ is at least $n! = (|\varphi| - 4)!$, and so exponential in the size of $\varphi$. 
\end{example}

\subsection{A polynomial reduction}
\label{subsec:poly}
As anticipated in the introduction, in order to avoid the exponential blowup illustrated by the previous example, we consider a second method that relies on finding an ordinary formula equivalid to the contextual formula. This will lead us to the complexity result of \cref{lem:bool:complexity}.

\begin{restatable}{proposition}{boolMethodTwo} \label{thm:bool:method2}
A propositional contextual formula $\varphi \in \boolc$ is valid if{}f the ordinary propositional formula
\begin{equation*}
\varphi_e \coloneq \left(\bigwedge_{c[\ctxarg_1], c[\ctxarg_2] \in \consub{\varphi}} (\decon{\ctxarg_1} \rightarrow \decon{\ctxarg_2}) \to (p_{\Context{c}{\ctxarg_1}} \rightarrow p_{\Context{c}{\ctxarg_2}}) \right) \to \decon{\varphi}
\end{equation*}
is valid.
\end{restatable}

\begin{sketch}
We follow here the same ideas of \cref{sec:bool:method1}. ($\Rightarrow$) If $\varphi_e$ is valid, for a given substitution $\sigma$ and Kripke structure $\mathcal{K}$, we define the Kripke structure $\mathcal{K}'$ of \cref{lem:bool:modelprime}. After showing again that $\feval[\valuation']{\decon{\phi}} = \feval{\sigma(\phi)}$ for every subformula, we see that the condition of $\varphi_e$ holds through a calculation, and then its conclusion yields $\feval[\valuation']{\decon{\varphi}} = \feval{\sigma(\varphi)} = 1$, so $\varphi$ is valid. ($\Leftarrow$) If $\varphi$ is valid, so is $\cninst_\varphi(\varphi)$ with a valuation $\valuation$. We show $\varphi_e$ holds under the same valuation. This is immediate if the premise does not hold. Otherwise, we can use the monotonicity encoded in the premise of $\varphi_e$ to almost repeat the calculation on $\cninst_\varphi(\Context{c}{\ctxarg})$ in \cref{lem:bool:modelprime} and conclude $\feval{\cninst_\varphi(\varphi)} = \feval[\valuation']{\decon{\varphi}} = 1$.
\end{sketch}

\begin{corollary} \label{lem:bool:complexity}
The validity and satisfiability problems for contextual propositional formulas are co-\textsc{np}-complete and \textsc{np}-complete, respectively.
\end{corollary}

\begin{proof}
\cref{thm:bool:method2} gives a polynomial reduction to validity of ordinary formulas.  Indeed,
$
	|\decon{\varphi}| \leq |\varphi| + |\varphi|^3 \cdot (2|\varphi| + 5) \leq 8 |\varphi|^4
$.
For satisfiability, it suffices to replace the top implication of $\varphi_e$ by a conjunction.
\end{proof}

\begin{remark}
In the propositional calculus, once we assign truth values to the atomic propositions every formula is equivalent to either $\true$ or $\false$. Similarly,  every context  is equivalent to  $\true$, $\false$, or $\hole$. Hence, an alternative method to check validity of a contextual propositional formula is to check the validity of all possible instantiations of the context variables with these three contexts. However, for $n$ different context variables, this requires $3^n$ validity checks.
\end{remark}
\begin{remark}
Other examples of valid identities are $\Context{c}{p \wedge q} \equiv \Context{c}{p} \wedge \Context{c}{q}$, $\Context{c}{p \vee q} \equiv \Context{c}{p} \vee \Context{c}{q}$, and $\Context{c}{p} \equiv \Context{c}{c[p}]$. Example of valid entailments are $(p \leftrightarrow q) \models (\Context{c}{p} \leftrightarrow \Context{c}{q})$ and $(p \to q) \models (\Context{c}{p} \to \Context{c}{q})$; the entailments in the other direction are not valid, as witnessed by the instantiation $\sigma(c) := \false$. Finally, $p \equiv \Context{c}{p}$ is an example of an identity that is not valid in any direction.
All these facts can be automatically checked using any of the methods described in the section.
\end{remark}

\section{Validity of contextual  \textmu-calculus formulas} \label{sec:mucalc}

We extend the reductions of \cref{sec:boolean}  to the contextual modal $\mu$-calculus. In particular, this requires introducing a new definition of canonical instantiation and a new equivalid formula. The main difference with the propositional case is that contexts may now contain free variables (that is, variables that are bound outside the context). For example, in the unfolding rule we find the context $\Context{c}{X}$, and $X$ appears free in the argument of $c$.  This problem will be solved by replacing each free variable $X$ by either the fixpoint subformula that binds it, or by a fresh atomic proposition $p_X$.  We will also need to tweak decontextualizations. More precisely, the canonical instantiation will have the shape
\[ \cninst_\varphi(\context) \coloneq \bigwedge_{\Context{c}{\ctxarg} \in \consub{\varphi}} \left(\AG (\hole \to \devarf{\ctxarg})\right)  \to \ctxap \]
\noindent where $\AG \psi$ is an abbreviation for $\nu X. ([\cdot] X \wedge \psi)$, and $\devarf{\ctxarg}$ is a slight generalization of $\decon{\ctxarg}$.

Throughout the section we let $\mucx$ and $\mucc$ denote the sets of all ordinary and contextual formulas of the contextual $\mu$-calculus over sets $\AP$, $V$, and $\contextVars$, of atomic propositions, variables, and context variables, respectively. Further, we assume w.l.o.g.\ that all occurrences of a variable in a formula are either bound or free, and that distinct fixpoint subformulas have distinct variables.  The following notation is also used throughout:

\begin{definition}
Given a formula $\varphi \in \mucc$ and a bound variable $X$ occurring in $\varphi$, we let $\alpha X. \varphi_X$ denote the unique fixpoint subformula of $\varphi$ binding $X$.
\end{definition}

\subsection{Variable and propositional substitutions}

A key tool to obtain the results of \cref{sec:boolean} was the substitution lemma for propositional logic. On top of atomic propositions, the $\mu$-calculus has also variables, and we need separate substitution lemmas for both of them.
We start with the variable substitution lemma. In this case, we have a $\mu$-calculus formula $\varphi$ with some free variables $X \in \FV{\varphi}$ and we want to replace them by closed formulas $\sigma(X) \in \mucx$. As usual, bound variables are not replaced by variable substitutions, i.e.\ $\sigma(\alpha X. \phi) = \alpha. \sigma|_{V \setminus \{X\}}(\phi)$. The following lemma is a direct translation of the substitution lemma for propositional logic.

\begin{restatable}[variable substitution lemma]{lemma}{subsmuvars} \label{lem:mucalc:subsvar}
For any Kripke structure with set of states $S$, valuation $\eta : V \to \mathcal{P}(S)$, and substitution $\sigma : V \to \mucx$ such that $\sigma(X)$ is either $X$ or a closed formula for all $X \in V$,
we have $\mueval[\eta]{\sigma(\varphi)} = \mueval[\eta']{\varphi}$ where $\eta'$ is defined by $\eta'(X) := \mueval[\eta]{\sigma(X)}$.
\end{restatable}

Replacing atomic propositions is more subtle, since they can be mapped to non-ground $\mu$-calculus formulas. While propositions have a fixed value,
the semantics of their replacements may depend on the valuation, which could be the dynamic result of fixpoint calculations appearing in the formula. Consequently, the substitution lemma will not work for an arbitrary valuation like in \cref{lem:mucalc:subsvar}, but only for those who \emph{match} the values of the fixpoint variables of the formula.

\begin{definition}[fixpoint valuation]
$\eta$ is a \emph{fixpoint valuation} of a formula $\varphi \in \mucx$ iff $\eta(X) = \mueval[\eta]{\alpha X. \phi}$ for every bound variable $X$ of $\varphi$.
\end{definition}

\begin{restatable}[propositional substitution lemma]{lemma}{subsmuprops} \label{lem:mucalc:subsap}
For any Kripke structure $\mathcal{K} = (S, {\to}, I, \AP, \ell)$, formula $\varphi \in \mucx$, substitution $\sigma : \AP \to \mucx$ such that $\sigma(p) = p$ if $p$ appears negated in $\varphi$, and valuation $\eta : V \to \mathcal P(S)$ such that $\eta(X) = \mueval{\alpha X.\sigma(\phi_X)}$ for every subformula $\alpha X.\phi_X$ of $\varphi$, we have $\mueval[\mathcal{K}, \eta]{\sigma(\varphi)} = \mueval[\mathcal{K'}, \eta]{\varphi}$ where $\mathcal{K}' = (S, {\to}, \AP, I, \ell')$ is the Kripke structure with $\ell'(p) = \mueval{\sigma(p)}$ for every $p \in \AP$.
\end{restatable}

These two lemmas will be very helpful in the proof of the main theorems, where we turn subformulas and variables into atomic propositions to make formulas like $\cninst_\varphi(\varphi)$ and $\varphi_e$ ordinary and closed, respectively. For example, consider $\mu X. \Context{c}{X} = \Context{c}{\mu X. \Context{c}{X}}$. If we take $\decon{\psi}$ instead of $\devarf{\psi}$ in the canonical instantiation, we  obtain
\begin{equation}
\cninst_\varphi(\context) = \left(\left(\AG (\hole \to {\color{red}X})\right) \to \ctxap{c}{X}\right) \wedge \left(\left(\AG (\hole \to \mu X. \ctxap{c}{X})\right) \to \ctxap{c}{\mu X. \Context{c}{X}}\right) \label{eq:xfree}
\end{equation}
Since $X$ is free in this context, which value should it take? The following lemma proves there is a unique fixpoint valuation $\mint$ for each formula $\varphi$ and Kripke structure $\mathcal{K}$. This will give the answer to this question.

\begin{restatable}[existence of fixpoint valuation]{lemma}{mucMint} \label{lem:mucalc:mint}
For every Kripke structure $\mathcal{K}$ and closed formula $\varphi \in \mucx$, there is a unique fixpoint valuation $\mint$, up to variables that do not appear in $\varphi$.
\end{restatable}

Moreover,  in (\ref{eq:xfree}), we will be interested in getting rid of the free occurrence of $X$ to reduce the problem to validity of ordinary closed formulas. The following lemma claims that we can replace the free variables in the formula by the fixpoint subformula defining those variables, without changing the semantics of the formula. We call the substitution achieving this, which we show to be independent of any Kripke structure, the \emph{fixpoint substitution} of $\varphi$.

\begin{restatable}[fixpoint substitution]{lemma}{mucMintsubs} \label{lem:mucalc:mintsubs}
For every closed formula $\varphi \in \mucx$, there is a (unique) variable substitution $\msigma : X \to \mucx$ such that $\msigma(X)$ is closed and $\msigma(X) = \msigma(\alpha X. \phi_X)$. Moreover, for every Kripke structure $\mathcal K$, every subformula $\phi$ of $\varphi$, and every fixpoint valuation $\eta$ for $\varphi$, we have $\mueval[]{\sigma(\phi)} = \mueval[\eta]{\phi}$.
\end{restatable}

\subsection{Canonical instantiation}
We are now ready to define the canonical instantiation of a contextual formula.

\begin{definition}[canonical instantiation of a contextual formula]
\label{def:canonicalinstantiation}
Let $\varphi \in \mucc$ be a contextual formula of the $\mu$-calculus. Given a subformula $\psi$ of $\varphi$, let $\devarf{\psi}$ be the result of
applying to $\decon{\psi}$ its fixpoint substitution.
The \emph{canonical instantiation} of $\varphi$ is the mapping $\cninst_\varphi : \contextVars \to \contextSet$ defined by
\[ \cninst_\varphi(\context) \coloneq \bigwedge_{\Context{c}{\ctxarg} \in \consub{\varphi}} \left(\AG (\hole \to \devarf{\ctxarg})\right)  \to \ctxap \]
where $\AG \psi$ is an abbreviation for $\nu X. ([\cdot] X \wedge \psi)$ for some fresh variable $X$.
\end{definition}

We prove that $\varphi$ is valid (resp.\ satisfiable) iff $\cninst_\varphi(\varphi)$ is valid (satisfiable). We need two lemmas. The first one is the extension of \cref{lem:bool:monotonic} to the $\mu$-calculus.

\begin{restatable}[monotonicity]{lemma}{mucGamma} \label{lem:mucalc:monotonic}
For every $\varphi, \psi, \psi' \in \mucx$, where only $\varphi$ may contain $\hole$, fixpoint valuation $\mint$, and every $s \in S$, if
$s \in \mueval{\AG (\psi \to \psi')}$,
then $s \in \mueval[\mint]{\varphi[\hole/\psi]}$ implies $s \in \mueval[\mint]{\varphi[\hole/\psi']}$.
\end{restatable}

The second lemma is the key one.

\begin{restatable}{lemma}{mucalcModelPrime} \label{lem:mucalc:modelprime}
For every $\varphi \in \mucx$, instantiation $\sigma$, and Kripke structure $\mathcal{K}= (S, \to, I, \AP, \ell)$, there is a Kripke structure $\mathcal{K}'=(S, \to, \AP', I, \ell')$, where $\AP \subseteq \AP'$ and $\ell'$ extends $\ell$, such that $\mueval[\mathcal{K}]{\sigma(\varphi)} = \mueval[\mathcal{K}']{\cninst_\varphi(\varphi)}$.
\end{restatable}

\begin{sketch}
The ideas of \cref{lem:bool:modelprime} are reproduced here, although with the additional complication of $\mu$-calculus variables. Given $\varphi$, $\sigma$, and $\mathcal{K}$, we define  $\mathcal{K}' \coloneq (S, {\to}, I, \AP', \ell')$, $\AP' \coloneq \AP \cup \{\ctxap \mid \Context{c}{\ctxarg} \in \consub{\varphi} \}$, $\ell'(p) = \ell(p)$ if $p \in \AP$, and $\ell'(\ctxap) = \mueval[\mint]{\sigma(\Context{\context}{\ctxarg})}$ using the fixpoint valuation $\mint$ of \cref{lem:mucalc:mint} for $\sigma(\varphi)$. First, we show $\mueval[\mint]{\devarf{\phi}} = \mueval[\mint]{\decon{\phi}}$ for every subformula $\phi$ of $\varphi$ using the variable substitution lemma (\cref{lem:mucalc:subsvar}) with $\gamma_*(X) = \alpha X. \varphi_X$. Then, like in the proposition case, we prove $\mueval[\mint]{\sigma(\phi)} = \mueval[\mint]{\decon{\phi}}$ for any subformula $\phi$ using the propositional substitution lemma (\cref{lem:mucalc:subsap}) with $\gamma_{\sigma}(\ctxap) = \sigma(\Context{c}{\ctxarg})$. Finally, we prove $\mueval[\mint]{\sigma(\phi)} = \mueval[\mint]{\cninst_\varphi(\phi)}$ by induction using some calculation (essentially the same in \cref{lem:bool:modelprime}) by the monotonicity of \cref{lem:mucalc:monotonic} on the expression of $\cninst_\varphi(c)$ as well as the propositional substitution lemma.
\end{sketch}

\begin{theorem} \label{thm:mucalc:main}
For every $\varphi \in \mucc$,$\varphi$ is valid (resp.\ satisfiable) iff $\cninst_\varphi(\varphi) \in \mucx$ is valid (satisfiable).
\end{theorem}

\begin{proof}
For validity: ($\Rightarrow$) If $\varphi$ is valid, then $\sigma(\varphi)$ is valid for every instantiation $\sigma$. In particular, $\cninst_\varphi(\varphi)$ is valid.
($\Leftarrow$) Assuming $\cninst_\varphi(\varphi)$ is valid and for any instantiation $\sigma$, we must prove that $\sigma(\varphi)$ is also valid. Let $\mathcal{K}$ be a Kripke structure, \cref{lem:mucalc:modelprime} claims there is another Kripke structure $\mathcal{K}'$ such that $\mueval[\mathcal{K}]{\sigma(\varphi)} = \mueval[\mathcal{K}']{\cninst_\varphi(\varphi)}$, so $\mathcal{K} \models \sigma(\varphi)$ iff $\mathcal{K}' \models \cninst_\varphi(\varphi)$. Moreover, we have $\mathcal{K}' \models \cninst_\varphi(\varphi)$ because $\cninst_\varphi(\varphi)$ is valid, so $\sigma(\varphi)$ is valid too since $\mathcal{K}$ is arbitrary.

For satisfiability, ($\Leftarrow$) If $\cninst_\varphi(\varphi)$ is satisfiable, then $\varphi$ is satisfiable by definition. ($\Rightarrow$) If $\varphi$ is satisfiable, then $\mathcal{K} \models \sigma(\varphi)$ for some $\sigma$ and $\mathcal{K}$. \Cref{lem:mucalc:modelprime} then ensures $\mathcal{K}' \models \cninst_\varphi(\varphi)$ for some $\mathcal{K'}$, so $\cninst_\varphi(\varphi)$ is satisfiable.
\end{proof}

As in the propositional case, the length of $\cninst_\varphi(\varphi)$ grows  exponentially in the nesting depth of the context. However, in the $\mu$-calculus  it can also grow exponentially even for non-nested contexts. The reason is that applying the  fixpoint substitution to a $\mu$-formula can yield an exponentially larger formula.

\begin{example} \label{example:subsblowup}
Consider $\varphi = \mu X_1. \cdots \mu X_n. X_1 \wedge \cdots \wedge X_n$ for any $n \in \N$, whose size is $|\varphi| = 3 n - 1$. It can be proven by induction that $|\sigma(X_k)| = 2^{k-1} (3n - 2) + 1$, so $|\sigma(X_n)| = 2^{n-1} (3n - 2) + 1 = 2^{\frac{1}{3}(|\varphi| - 2)} (|\varphi| + 1) + 1 \in O(2^{|\varphi|})$.\footnote{The difficulty in the previous example are fixpoint formulas with free variables. Otherwise, $|\sigma(X)| = |\sigma(\alpha X. \varphi_{X})| = |\alpha X. \varphi_{X}| \leq |\varphi|$.}
\end{example}

Both problems are solved in the next section by giving an alternative reduction to validity/satisfiability of ordinary $\mu$-calculus formulas.

\subsection{A polynomial reduction}
Given a contextual formula $\varphi$ of the $\mu$-calculus, we construct an ordinary formula $\varphi_e$ equivalid to $\varphi$ of polynomial size in $\varphi$. Following the idea of \cref{thm:bool:method2}, we replace all context occurrences in $\varphi$ by fresh atomic propositions, and insert additional conditions to ensure that the values of these propositions are consistent with what they represent. However, in the $\mu$-calculus, these conditions may introduce unbound variables, and the strategy to remove them in \cref{thm:mucalc:main} involves the exponential blowup attested by \cref{example:subsblowup}. To solve this problem, we also replace the free $\mu$-calculus variables in the context occurrences by fresh atomic propositions; further, we add additional clauses to ensure that they take a value consistent with the fixpoint calculation.

\begin{definition}[equivalid formula]
For every contextual formula $\varphi \in \mucc$,
\begin{itemize}
	\item let $F = \bigcup_{\Context{c}{\ctxarg} \in \consub{\varphi}} \FV{\ctxarg}$ be the free variables in all context arguments of $\varphi$;
	\item for every $X \in F$, let $p_{X}$ be a fresh variable that does not occur in $\varphi$; and
	\item for every subformula $\phi$ of $\varphi$, let $\devarp{\phi}$ be the result of replacing every free occurrence of $X$  in the formula $\decon{\phi}$ by $p_{X}$.
	\end{itemize}
We define the ordinary formula $\varphi_e \in \mucx$ as
\begin{equation*}
\kern-0.5ex\left(\bigwedge_{\substack{\Context{c}{\psi_1}, \Context{c}{\psi_2} \\ \in \consub{\varphi}}} \kern-1.5ex \AG (\AG (\devarp{\ctxarg_1} \to \devarp{\ctxarg_2}) \to (\ctxap{c}{\ctxarg_1} \to \ctxap{c}{\ctxarg_2})) \wedge \bigwedge_{X \in F} \AG (p_{X} \leftrightarrow \alpha X. \phi_{X}^+) \right) \to \devarp{\varphi}
\end{equation*}
\end{definition}

We say $\mathcal{K'}$ is an extension of $\mathcal{K}$ if $\mathcal{K}' = (S, \{\to\}, I, \AP', \ell')$, $\AP \subseteq \AP'$, and $\ell'|_{\AP} = \ell$.

\begin{restatable}{proposition}{mucalcModelsTwo} \label{thm:mucalc:models2}
For every contextual formula $\varphi \in \mucc$, instantiation $\sigma$, and Kripke structure $\mathcal{K}=(S, {\to}, I, \AP, \ell)$,
\begin{enumerate}
	\item there is an extension $\mathcal{K'}$ of $\mathcal{K}$ such that $\mathcal{K} \models \sigma(\varphi)$ iff $\mathcal{K}' \models \varphi_e$.
	\item for any extension $\mathcal{K}'$ of $\mathcal{K}$ such that $\mathcal{K}'$ is a model for the premise of $\varphi_e$, $\mathcal{K}' \models \cninst_\varphi(\varphi)$ iff $\mathcal{K}' \models \varphi_e$.
\end{enumerate}
\end{restatable}

\begin{sketch}
We follow the main strategy of \cref{thm:bool:method2}. (1) Using the valuation $\mint$ given by \cref{lem:mucalc:mint} for $\sigma(\varphi)$, we define the Kripke structure $\mathcal{K}'$ of \cref{lem:mucalc:modelprime} with some more variables $\ell'(p_X) = \mueval[\mathcal{K}, \mint]{\alpha X. \varphi_X}$. Again, for every subformula, we prove $\mueval[\mathcal{K}', \mint]{\decon{\phi}} = \mueval[\mathcal{K}', \mint]{\devarp{\phi}}$, $\mueval[\mathcal{K}', \mint]{\decon{\phi}} = \mueval[\mathcal{K}, \mint]{\sigma(\phi)}$ invoking the appropriate substitution lemmas. This let us prove that the premise of $\varphi_e$ holds because of the monotonicity of contexts reflected in \cref{lem:mucalc:monotonic} and the matching definitions of $\mint$ and $\ell'(p_X)$. Hence, $\varphi_e$ is equivalent to its conclusion, and we have proven $\mueval[\mathcal{K}', \mint]{\decon{\varphi}} = \mueval[\mathcal{K}, \mint]{\sigma(\varphi)}$, which implies the statement.

(2) Let $\varphi_e = \varphi_p \to \decon{\varphi}$, we now assume $\mathcal{K}' \models \varphi_p$ and must prove $\mathcal{K}' \models \cninst_\varphi(\varphi)$ iff $\mathcal{K}' \models \varphi_e$, or equivalently iff $\mathcal{K}' \models \devarp{\varphi}$. Using the fixpoint valuation $\mint$ of $\cninst_\varphi(\varphi)$, we inductively extract from the premise that $\ell'(p_X) = \mint(X)$ and $\ell'(\ctxap) = \mueval[\mint]{\cninst_\varphi(\Context{c}{\ctxarg})}$ with the usual calculations and substitutions. Then, we conclude that the right argument of the top implication in $\varphi_e$ satisfies $\mueval[]{\devarp{\varphi}} = \mueval[]{\sigma(\varphi)}$, which implies the statement.
\end{sketch}

\begin{restatable}{theorem}{mucalcMethodTwo} \label{thm:mucalc:method2}
A contextual  $\mu$-calculus formula $\varphi \in \mucc$ is valid iff $\varphi_e \in \mucx$ is valid.
\end{restatable}

\begin{proof}
($\Leftarrow$) $\varphi$ is valid if $\sigma(\varphi)$ is valid for every instantiation $\sigma$. Let $\mathcal{K}$ be any Kripke structure, $\mathcal{K} \models \sigma(\varphi)$ must hold. However, \cref{thm:mucalc:models2} ensures there exists $\mathcal{K}'$ such that $\mathcal{K} \models \sigma(\varphi)$ if $\mathcal{K} \models \varphi_e$. Since $\varphi_e$ is valid, we are done.
($\Rightarrow$) If $\varphi$ is valid, so is $\cninst_\varphi(\varphi)$. We should prove that $\varphi_e$ is valid, which means $\mathcal{K} \models \varphi_e$ for all $\mathcal{K}$. If the premise of $\varphi_e$ does not hold, $\mathcal{K} \models \varphi_e$ trivially. Otherwise, \cref{thm:mucalc:models2} reduce the problem to $\mathcal{K} \models \cninst_\varphi(\varphi)$, which holds by hypothesis.
\end{proof}

\begin{corollary}
The validity and satisfiability problems for contextual $\mu$-calculus formulas are \textsc{exptime}-complete.
\end{corollary}

\begin{proof}
The validity and satisfiability problems for ordinary formulas of the $\mu$-calculus are \textsc{exptime}-complete~\cite{handbookMucalc}. \cref{thm:mucalc:method2} gives a polynomial reduction from contextual to ordinary validity. Indeed, a rough bound is
$
|\varphi_e| \leq |\varphi|^3 \cdot (2 |\varphi| + 5) + |\varphi| \cdot (|\varphi| + 2) + |\varphi| \leq 11 |\varphi|^4
$. For satisfiability, we can again replace the top implication of $\varphi_e$ by a conjunction.
\end{proof}

\subsection{Validity of contextual  CTL formulas} \label{sec:ctl}
We assume that the reader is familiar with the syntax and semantics of CTL (see e.g.\ \cite{ClarkeHV18}), and only fix a few notations.
The syntax of contextual CTL over a set $\AP$ of atomic propositions is:
\[ \varphi \Coloneq p \mid \neg\, p \mid  \varphi \wedge \varphi \mid \varphi \vee \varphi \mid \Context{c}{\varphi} \mid
	  \A (\varphi \U \varphi) \mid  \A (\varphi \W \varphi) \mid \mathbf{E}\, (\varphi \U \varphi) \mid \mathbf{E}\, (\varphi \W \varphi) \label{eq:CTLsyntax}
\]
where $p  \in \AP$, $c \in C$, and $\mathbf{U}$, $\mathbf{W}$ are the strong until and weak until operators. As for the $\mu$-calculus, the syntax of contextual CTL-formulas adds a term $\Context{c}{\varphi}$, and the syntax of CTL contexts adds the hole term $\hole$.

Given a Kripke structure $\mathcal{K} = (S, {\to}, I, \AP, \ell)$ and a mapping $\ell \colon S \to \mathcal{P}(S)$, the semantics assigns to each ordinary formula $\varphi$ the set $\mueval[]{\varphi} \subseteq S$ of states satisfying $\varphi$. For example, $\mueval[]{\mathbf{E}\, (\varphi_1 \W \varphi_2)}$ is the set of states $s_0$ such that some infinite  path $s_0 s_1 s_2 \cdots$ of the Kripke structure satisfies either  $s_i \in \mueval[]{\varphi_1}$ for every $i \in \mathbb{N}$, or $s_k \in \mueval[]{\varphi_2}$ and $s_0, \ldots, s_{k-1} \in \mueval[]{\varphi_1}$ for some $k \geq 1$. We extend the semantics to contexts and contextual formulas as for the $\mu$-calculus.

We proceed to solve the validity problem for contextual CTL using the syntax-guided translation from CTL to the $\mu$-calculus~\cite[Pag.\ 1066]{ctltrans}. The translation
assigns to each CTL-formula $\varphi$ a closed formula $\ctltomc{\varphi}$ of the $\mu$-calculus such that $\mueval[]{\varphi}=\mueval[]{\ctltomc{\varphi}}$ holds for every 
Kripke structure $\mathcal{K}$ and mapping $\ell$.

\begin{definition}[CTL to $\mu$-calculus translation]
For any context or contextual CTL formula $\varphi$ we inductively define the $\mu$-calculus formula $\ctltomc{\varphi}$ by
\begin{multicols}{2}
\begin{enumerate}
\item $\ctltomc{p} = p$
\item $\ctltomc{(\neg\, p)} = \neg\, p$
\item $\ctltomc{(\hole)} = \hole$
\item $\ctltomc{(\Context{c}{\ctxarg})} = \Context{c}{\ctltomc{\ctxarg}}$
\item $\ctltomc{(\varphi_1 \wedge \varphi_2)} = \ctltomc{\varphi_1} \wedge \ctltomc{\varphi_2}$
\item $\ctltomc{(\varphi_1 \vee \varphi_2)} = \ctltomc{\varphi_1} \vee \ctltomc{\varphi_2}$
\item $\ctltomc{\A (\varphi_1 \U \varphi_2)} = \mu X. (\mcAX X \wedge \ctltomc{\varphi_1}) \vee \ctltomc{\varphi_2}$.
\item $\ctltomc{\A (\varphi_1 \W \varphi_2)} = \nu X. (\mcAX X \wedge \ctltomc{\varphi_1}) \vee \ctltomc{\varphi_2}$.
\item $\ctltomc{\mathbf{E}\, (\varphi_1 \U \varphi_2)} = \mu X. (\mcEX X \wedge \ctltomc{\varphi_1}) \vee \ctltomc{\varphi_2}$.
\item $\ctltomc{\mathbf{E}\, (\varphi_1 \W \varphi_2)} = \nu X. (\mcEX X \wedge \ctltomc{\varphi_1}) \vee \ctltomc{\varphi_2}$.
\end{enumerate}
\end{multicols}
\noindent The translations of $\mathbf{AF}$, $\mathbf{AG}$, $\mathbf{EF}$, and $\mathbf{EG}$ follow by instantiating (7-10) appropriately.
\end{definition}

The proof of the following corollary can be found in \fullversionref. Intuitively, it is a consequence of the fact that the canonical instantiation of
Definition \ref{def:canonicalinstantiation} is a formula of the CTL-fragment of the $\mu$-calculus.

\begin{restatable}{corollary}{ctlMain} \label{thm:ctl:main}
For any contextual CTL formula $\varphi$, let $\cninst_\varphi : \contextVars \to \mathrm{CTL}$ be the instantiation of contexts defined by
\[ \cninst_\varphi(c) \coloneq \bigwedge_{\Context{c}{\ctxarg} \in \consub{\varphi}} \left(\AG (\hole \to \decon{\ctxarg})\right) \to \ctxap \]
Then, the following statements are equivalent:
\begin{enumerate}
	\item $\varphi$ is valid,
	\item $\cninst_\varphi(\varphi)$ is valid, and
	\item $\varphi_e \coloneq \left(\bigwedge_{\Context{c}{\ctxarg_1}, \Context{c}{\ctxarg_2} \in \consub{\varphi}} \AG (\AG (\decon{\ctxarg_1} \to \decon{\ctxarg_2}) \to (\ctxap{c}{\ctxarg_1} \to \ctxap{c}{\ctxarg_2}))\right) \to \decon{\varphi}$ is valid.\label{thm:ctl:main:method2}
\end{enumerate}
\end{restatable}

\begin{sketch}
A straightforward induction shows that $\ctltomc{(\sigma(\varphi))} = \ctltomc{\sigma}(\ctltomc{\varphi})$ for any instantiation $\sigma$. Moreover, $\ctltomc{\cninst_{\varphi}} = \cninst_{\ctltomc{\varphi}}$ and $\ctltomc{\varphi_e} = (\ctltomc{\varphi})_e$. Hence, going back and forth between CTL and the $\mu$-calculus, we can translate \cref{thm:mucalc:main,thm:mucalc:method2} to CTL.
\end{sketch}

\begin{corollary}
The validity problem for contextual CTL formulas is \textsc{exptime}-complete.
\end{corollary}

\begin{proof}
The validity problem for CTL is known to be \textsc{exptime}-complete~\cite{ctlSatisfiability}, it is a specific case of the contextual validity problem, and \cref{thm:ctl:main} gives a polynomial reduction from the latter to the former.
\end{proof}

\begin{example}
Let us use item (2) of \cref{thm:ctl:main} to show that the Boole-Shannon expansion  is not valid in CTL. As in \cref{ex:shannon2}, let $\varphi:= \Context{c}{p} \leftrightarrow (p \wedge \Context{c}{\true} ) \vee (\neg\, p \wedge \Context{c}{\false})$. By definition, $\ \cninst_\varphi(\varphi)  :=  \cninst_\varphi(\Context{c}{p})  \leftrightarrow  \big( (p \wedge  \cninst_\varphi(\Context{c}{\true})) \vee (\neg\, p \wedge  \cninst_\varphi(\Context{c}{\false})) \big)$. Simplification yields
 $$\begin{array}{rcc}
& &    p_{\Context{c}{p}}    \wedge     p_{\Context{c}{\true}}   \wedge     (\AG \neg p \to p_{\Context{c}{\false}})    \\
\cninst_\varphi(\varphi)  & \equiv & \leftrightarrow \\
& &  (p \wedge (\AG p \to p_{\Context{c}{p}}) \wedge p_{\Context{c}{\true}}) \vee
 (\neg p \wedge p_{\Context{c}{p}} \wedge p_{\Context{c}{\true}} \wedge p_{\Context{c}{\false}})
\end{array}$$
This formula is not valid. For example, take any Kripke structure with a state $s$ satisfying $p$ and $p_{\Context{c}{\true}}$, but neither $p_{\Context{c}{p}}$ nor $\AG p$.
Then $s$ satisfies the right-hand side of the bi-implication, because it satisfies the left disjunct, but not the left-hand side, because it does not satisfy $p_{\Context{c}{p}}$. By \cref{thm:ctl:main}, the Boole-Shannon expansion is not valid.
\end{example}

Either item (2) or (3) of \cref{thm:ctl:main} can be used to check, for instance, that the substitution rules $\AG (a \leftrightarrow b) \models \AG (\Context{c}{a} \leftrightarrow \Context{c}{b})$, and $\AG (a \to b) \models \AG (\Context{c}{a} \to \Context{c}{b})$ do hold.

\subsection{Validity of contextual LTL formulas} \label{sec:ltl}

The syntax of LTL is obtained by dropping $\mathbf{E}$ and $\mathbf{A}$ from the syntax of CTL. Formulas are interpreted over infinite sequences of atomic propositions \cite{ClarkeHV18},
and a state $s_0$ of a Kripke structure satisfies a formula $\varphi$ if every infinite path $s_0 s_1 s_2 \cdots$ of the Kripke structure satisfies $\varphi$. The Kripke structure itself satisfies $\varphi$ if all its initial states satisfy $\varphi$. 

Unlike for CTL, there is no syntax-guided translation from LTL to $\mu$-calculus.
However, there is one for \emph{lassos}, finite Kripke structures in which every state has exactly one infinite path rooted at it. 

\begin{definition}
A Kripke structure $\mathcal{K}=(S, {\to}, I, \AP, \ell)$ is a \emph{lasso} if $S$ is finite and for every $s \in S$ there is exactly one state $s' \in S$ such that $s \to s'$.
\end{definition}

Consider the variation of the translation $\ctltomc{\varphi}$ from CTL where $\mathbf{X}$, $\mathbf{U}$, $\mathbf{G}$, and $\mathbf{F}$ are translated as $\mathbf{AX}$, $\mathbf{AU}$, $\mathbf{AG}$, and $\mathbf{AF}$. 
For example, we define $\ctltomc{(\varphi_1 \U \varphi_2)} := \mu X. (\mcAX X \wedge \ctltomc{\varphi_1}) \vee \ctltomc{\varphi_2}$. We have:

\begin{restatable}{lemma}{lemLasso} \label{lem:lasso}
An LTL formula is valid if{}f it holds over all lassos. Further, for every lasso $\mathcal{K}$ and every formula $\varphi$ of LTL, $\mathcal{K} \models_{\text{LTL}} \varphi$ if{}f $\mathcal{K} \models \ctltomc{\varphi}$.
\end{restatable}

The results of \cref{thm:mucalc:main,thm:mucalc:method2} can be extended to LTL similarly to the CTL case. However, since $\ctltomc{\varphi}$ and $\varphi$ are only guaranteed to be equivalent in lassos, some care should be taken to always use them.

\begin{restatable}{corollary}{ltlMain} \label{thm:ltl:main}
For any LTL formula $\varphi$, let $\cninst_\varphi : \contextVars \to \mathrm{LTL}$ be the instantiation of contexts defined by
\[ \cninst_\varphi(c) \coloneq \bigwedge_{\Context{c}{\ctxarg} \in \consub{\varphi}} \left(\G (\hole \to \decon{\ctxarg})\right) \to \ctxap \]
Then, the following statements are equivalent
\begin{enumerate}
	\item $\varphi$ is valid,
	\item $\cninst_\varphi(\varphi)$ is valid, and \label{thm:ltl:main:method1}
	\item $\varphi_e \coloneq \left(\bigwedge_{\Context{c}{\ctxarg_1}, \Context{c}{\ctxarg_2} \in \consub{\varphi}} \G (\G (\decon{\ctxarg_1} \to \decon{\ctxarg_2}) \to (\ctxap{c}{\ctxarg_1} \to \ctxap{c}{\ctxarg_2}))\right) \to \decon{\varphi}$ is valid. \label{thm:ltl:main:method2}
\end{enumerate}
\end{restatable}

\begin{figure} \centering
\begin{tikzpicture}[>=latex, x=6em, y=3.5em,
    imp12/.style={},
    imp13/.style={decoration={snake, amplitude=.8pt, segment length=5pt, post length=3.5pt}, decorate, red},
    imp31/.style={dashed, blue},
]
	\node (CC) at (0, 1) {$\varphi$ is valid};
	\node (C) at (1, 1) {$\mathcal{L} \models \sigma(\varphi)$};
\node (M) at (1, 0) {$\mathcal{L} \models \ctltomc{\sigma}(\ctltomc{\varphi})$};
	\node (XC) at (-1, 1) {$\mathcal{L}' \models \cninst_\varphi(\varphi)$};
	\node (XM) at (-1, 0) {$\mathcal{L}' \models \cninst_{\ctltomc{\varphi}}(\ctltomc{\varphi})$};
	\node (EC) at (3, 1) {$\mathcal{L}' \models \varphi_e$};
	\node (EM) at (3, 0) {$\mathcal{L}' \models \ctltomc{\varphi_e}$};

	\node at (-2.25, 1) {LTL};
	\node at (-2.25, 0) {$\mu$-calculus};

\draw[->, imp12, transform canvas={shift={(0, -2pt)}}] (CC) -- (C);
	\draw[->, imp13, transform canvas={shift={(0, 2pt)}}] (CC) -- (C);
	\draw[->, imp12, transform canvas={shift={(-2pt, 0)}}] (C) -- (M);
	\draw[->, imp13, transform canvas={shift={(2pt, 0)}}] (C) -- (M);
	\draw[->, imp12] (M) edge node[below] {\scriptsize Lem.\ \ref{lem:mucalc:modelprime}} (XM);
    \draw[->] (M) edge[imp13] node[below, color=black] {\scriptsize Prop.\ \ref{thm:mucalc:models2}} (EM);
\draw[->, imp13, transform canvas={shift={(2pt, 0)}}] (EM) -- (EC);
    \draw[->, imp31, transform canvas={shift={(-2pt, 0)}}] (EC) -- (EM);
\draw[->, imp12, transform canvas={shift={(-2pt, 0)}}] (XM) -- (XC);
    \draw[->, imp31, transform canvas={shift={(2pt, 0)}}] (XM) -- (XC);

    \draw[->, imp31] (EM) -- ($(EM.south) + (0, -.3)$) -- node[below, color=black] {\scriptsize Prop.\ \ref{thm:mucalc:models2}} ($(XM.south) + (0, -.3)$) -- (XM.south);

\begin{scope}[shift={(0, -1.5)}]
        \node[font=\small] (I12) at (-2, 0) {$(2) \Rightarrow (1)$};
        \node[font=\small] (I13) at (0, 0) {$(3) \Rightarrow (1)$};
        \node[font=\small] (I32) at (2, 0) {$(2) \Rightarrow (3)$};
        \draw[imp12, line width=.7pt] ($(I12.east) + (1em, 0)$) -- ($(I12.east) + (4em, 0)$);
        \draw[imp13, line width=.7pt, decoration={post length=0pt}] ($(I13.east) + (1em, 0)$) -- ($(I13.east) + (4em, 0)$);
        \draw[imp31, line width=.7pt] ($(I32.east) + (1em, 0)$) -- ($(I32.east) + (4em, 0)$);
    \end{scope}
\end{tikzpicture}
\caption{Proof summary of \cref{thm:ltl:main} (arrow is problem reduction).} \label{fig:ltl:main}
\end{figure}

\begin{sketch}
Like for CTL, $\ctltomc{(\sigma(\varphi))} = \ctltomc{\sigma}(\ctltomc{\sigma})$, $\ctltomc{\cninst_\varphi} = \cninst_{\ctltomc{\varphi}}$, and $\ctltomc{\varphi_e} = (\ctltomc{\varphi})_e$. Each implication of the equivalence can be derived as depicted in \cref{fig:ltl:main}. Notice that, while $\sigma(\varphi)$ and $\ctltomc{\sigma}(\ctltomc{\varphi})$ are not equivalent in general, they are when evaluated on a lasso by \cref{lem:lasso}. This allows going back and forth between LTL and the $\mu$-calculus, and \cref{lem:mucalc:modelprime,thm:mucalc:models2} complete the proof.
\end{sketch}

Using the procedure just described, we can check that the rules in \cref{fig:rewrite} are valid, that the Boole-Shannon expansion does not hold in LTL, and  one-side implications like $\Context{c}{\G p} \models \Context{c}{p}$, $\G (a \leftrightarrow b) \models \G (\Context{\varphi}{a} \leftrightarrow \Context{\varphi}{b})$ and $\G (a \to b) \models \G (\Context{\varphi}{a} \to \Context{\varphi}{b})$.

\section{Experiments} \label{sec:experiments}

We have implemented the methods of \cref{thm:bool:main,thm:bool:method2,thm:ctl:main,thm:ltl:main} in a prototype that takes two contextual formulas as input and tells whether they are equivalent, one implies another, or they are incomparable.\footnote{The prototype and its source code are publicly available at \url{https://github.com/ningit/ctxform}.} The prototype is written in Python and calls external tools for checking validity of ordinary formulas: MiniSat~\cite{minisat} through PySAT~\cite{pysat} for propositional logic, \spotv~\cite{spot} for LTL, and CTL-SAT~\cite{ctlsat} for CTL. No tool has been found for deciding $\mu$-calculus satisfiability, so this logic is currently not supported.

Most formulas of \cref{fig:rewrite} are solved in less than 10 milliseconds by the first and second methods. The hardest formula is the second one: using the equivalid formula, the largest automaton that appears in the process has 130 states and it is solved in 16.21 ms; with the canonical instantiation, the numbers are 36 and 20.22 ms. We have also applied small mutations to the rules of \cref{fig:rewrite} to yield other identities that may or may not hold (see \appendixref{sec:mutated}{C} for a list). Solving them takes roughly the same time and memory as the original ones. For CTL, the behavior even with small formulas is much worse because of the worse performance of CTL-SAT. The canonical context method takes 20 minutes to solve $\Context{c}{a \wedge b} \equiv \Context{c}{a} \wedge \Context{c}{b}$, while the method by the equivalid formula runs out of memory with that example and requires 22 minutes for the Boole-Shannon expansion. Hence, we have not continued with further benchmarks on CTL. The first three rows of \cref{fig:experiments} show the time, peak memory usage (in megabytes), and number of states of the automata (for LTL) required for checking the aforementioned examples. The experiments have been run under Linux in an Intel Xeon Silver 4216 machine limited to 8 Gb of RAM. Memory usage is as reported by Linux cgroups' memory controller.

In addition to these natural formulas, we have tried with some artificial ones with greater sizes and nested contexts to challenge the performance of the algorithm. We have considered two repetitive expansions of the rules in \cref{fig:rewrite}:
\begin{enumerate}
	\item There is a dual of the first rule in \cref{fig:rewrite} that removes a $\mathbf{W}$-node below a $\mathbf{U}$-node:
\[ \varphi \U \Context{c}{\psi_1 \W \psi_2} \equiv \varphi \U \Context{c}{\psi_1 \U \psi_2} \vee (\FG \varphi \wedge (\varphi \wedge \F \Context{c}{\true})  \W \Context{c}{\psi_1 \W \psi_2}).\]
Then, we can build formulas like $\Context{c_1}{\psi_1 \U \psi_2}\W \varphi$ ($n=1$), $\Context{c_1}{\psi_1 \U \Context{c_2}{\psi_2 \W \psi_3}} \W \varphi$ ($n=2$), $\Context{c_1}{\psi_1 \U \Context{c_2}{\psi_2 \W \Context{c_3}{\psi_3 \U \psi_4}}} \W \varphi$ ($n=3$), and so on, and apply  the first rule of the rewrite system and its dual to obtain the normalized right-hand side. \Cref{fig:experiments} shows that the first method does not finish within an hour for $n=2$, and the second reaches this time limit for $n=3$. For $n=2$ the second method checks the emptiness of an automaton of 1560 states (and another of 720 states for the other side of the implication).
	\item The third and fourth rules of \cref{fig:rewrite} can also be nested. We can consider $\Context{c_0}{\FG \Context{c_1}{p}}$ ($n=1$), $\Context{c_0}{\FG \Context{c_1}{\GF \Context{c_2}{p}}}$ ($n=2$), $\Context{c_0}{\FG \Context{c_1}{\GF \Context{c_2}{\FG \Context{c_3}{p}}}}$ ($n=3$), and so on. We also take $c_0 = c_1 = \cdots = c_n$ to make the problem harder.
As shown in \cref{fig:experiments}, we can solve up to $n=3$ within the memory constraints.
\end{enumerate}

\begin{table}
\centering\small
\begin{tblr}{
	colspec={Q[c,3.5em]Q[c,5ex]rrrrrr},
	vline{2-Y} = {2-Z}{solid},
	vline{3,6} = {2}{2-Z}{solid},
	hline{2}={3-Z}{solid},
	hline{3,5,6,7,10,Z}={solid}
}
	\SetCell[c=2]{r}                    &   & \SetCell[c=3]{c} Method 1 & & & \SetCell[c=3]{c} Method 2 & &   \\
	\SetCell[c=2]{r} Example            &   & Time        & Memory & States & Time      & Memory  & States \\
	\SetCell[r=2]{c} Shannon    & Bool      & 8.18 ms    & 3.67      &        & 15.08 ms  & 3.67      &        \\
	                            & LTL       & 5 ms    & 2.62       & 9       & 8.07 ms  & 2.62       & 12       \\
\SetCell[c=2]{r} Rules~\cite{jacm} (max)		&        & 23.88 ms    & 5.24       & 36     & 16.21 ms  & 4.71       & 130    \\
	\SetCell[c=2]{r} Mutated (max)		&        & 44.12 ms    & 5.72       & 48     & 31.68 ms  & 4.98       & 130    \\
	\SetCell[r=3]{c} (1)        & 0 & 1.66 ms     & 1.05       & 4      & 1.53 ms   & 1.05       & 4      \\
			                    & 1 & 17.40 ms    & 4.92       & 36     & 15.62 ms  & 4.46       & 130    \\
			                    & 2 & \textcolor{red}{timeout}     &          &        & 3:25 min  & 413.83      & 1560   \\
\SetCell[r=3]{c} (2)        & 1 & 36.19 ms    & 5.24      & 45     & 22.40 ms  & 5.24      & 80     \\
			                    & 2 & 623.07 ms   & 26.96     & 168    & 117.79 ms & 10.49      & 160    \\
			                    & 3 & 5:21 min    & 1245.95   & 1140    & 26.04 s   & 100.25      & 220    \\
\end{tblr}
\medskip
\caption{Compared performance with the challenging examples (memory in Mb).} \label{fig:experiments}
\end{table}

\section{Conclusions} \label{sec:conclusions}

We have presented two different methods to decide the validity and satisfiability of contextual formulas in propositional logic, LTL, CTL, and the $\mu$-calculus. Moreover, we have shown that these problems have the same complexity for contextual and ordinary formulas. Interesting contextual equivalences can now be checked automatically. In particular, we have replaced the manual proofs of the several LTL simplification rules in~\cite{jacm} to a few milliseconds of automated check.

While we have limited our exposition to formulas in negation normal form, and hence to monotonic contexts, the results for propositional logic, CTL, and LTL can be generalized to the unrestricted syntax of the corresponding logics and to arbitrary contexts. Some clues are given in \appendixref{sec:extensions}{B}.

\ifarxiv
\appendix

\section{Extended definitions and proofs} \label{sec:proofs}
Let $\mucx$ and $\mathbb{C}$ be the sets of ordinary and contextual formulas of the $\mu$-calculus or its sublogics. Let $\contextVars$ and $\contextSet$ be the set of context variables and contexts.

\subsection{Propositional logic}

\boolModelprime*

\begin{proof}
Consider $\valuation'(\ctxap) = \feval{\sigma(\Context{c}{\ctxarg})}$ for every $\Context{c}{\ctxarg} \in \consub{\varphi}$ and $\valuation'(x) = \valuation(x)$ otherwise. Since $\ctxap$ does not occur in $\sigma(\varphi)$, $\feval[\valuation']{\sigma(\varphi)} = \feval[\valuation]{\sigma(\varphi)}$, so we assume without loss of generality that $\valuation = \valuation'$ and drop the prime for the rest of the proof. First, observe the following claim:

\begin{claim} \label{clm:bool:arg}
$\feval{\sigma(\phi)} = \feval{\decon{\phi}}$ for any subformula $\phi$ of $\varphi$.
\end{claim}

Consider the substitution $\gamma$ such that $\gamma(\ctxap) = \sigma(\Context{c}{\ctxarg})$ for all $\Context{c}{\ctxarg} \in \consub{\varphi}$, and $\gamma(p) = p$ otherwise. It satisfies $\gamma(\decon{\phi}) = \sigma(\phi)$. Then, by the substitution lemma (\cref{lem:bool:subs}), $\feval{\gamma(\decon{\phi})} = \feval[\valuation']{\decon{\phi}}$ holds for a $\valuation'$ that only differs from $\valuation$ on the substituted variables. However, $\valuation'(\ctxap) = \feval{\gamma(\ctxap)} = \feval{\sigma(\Context{c}{\ctxarg})} = \valuation(\ctxap)$ by definition. Hence, $\valuation' = \valuation$, and $\feval{\sigma(\phi)} = \feval{\decon{\phi}}$.

\begin{claim} \label{clm:bool:final}
$\feval{\sigma(\phi)} = \feval{\cninst_\varphi(\phi)}$ for any subformula $\phi$ of $\varphi$.
\end{claim}

We proceed by induction on $\phi$. In case $\phi$ does not contain any context variable, $\sigma(\phi) = \phi = \cninst_\varphi(\phi)$, so the statement is trivially satisfied. The cases of conjunction and disjunction are also straightforward, because both instantiation and evaluation are defined homomorphically on the formulas. The remaining and most interesting case is that of formulas $\Context{c}{\ctxarg}$. On the one hand,
\begin{align*}
	\textstyle \feval{\cninst_\varphi(\Context{c}{\ctxarg})} &= \textstyle \feval{\bigwedge_{\Context{c}{\otherarg} \in \consub{\varphi}} (\cninst_\varphi(\ctxarg) \rightarrow \decon{\otherarg}) \to \ctxap{c}{\otherarg}} \\
	&= \textstyle \bigwedge_{\Context{c}{\otherarg} \in \consub{\varphi}} (\feval{\cninst_\varphi(\ctxarg)} \rightarrow \feval{\decon{\otherarg}}) \to \feval{\ctxap{c}{\otherarg}} \\
	&= \textstyle \bigwedge_{\Context{c}{\otherarg} \in \consub{\varphi}} (\feval{\sigma(\ctxarg)} \rightarrow \feval{\sigma(\otherarg)}) \to \feval{\sigma(\Context{c}{\otherarg})}
\end{align*}
where the last step holds by induction hypothesis on $\psi$ (which is a subformula of $\Context{c}{\psi}$) and \cref{clm:bool:arg} on $\otherarg$ (which may not be a subformula). For each conjunct, we consider two cases:
\begin{itemize}
	\item If $\feval{\sigma(\ctxarg)} \rightarrow \feval{\sigma(\otherarg)}$ does not hold, the clause is trivially true. Hence, we can ignore it in the conjunction.
	\item Otherwise, the clause takes the value $\feval{\sigma(\Context{c}{\otherarg})} = \feval{\sigma(c)[\hole/\sigma(\otherarg)])}$. Moreover, since the premise holds, we have $\feval{\sigma(\Context{c}{\ctxarg})} \to \feval{\sigma(\Context{c}{\otherarg})}$ by \cref{lem:bool:monotonic}.
\end{itemize}
One of the clauses is equal to $\feval{\sigma(\Context{c}{\ctxarg})}$ (when $\otherarg = \ctxarg$) and all others are either that value or true. Hence, $\feval{\cninst_\varphi(\Context{c}{\ctxarg})} = \feval{\sigma(\Context{c}{\ctxarg})}$, and we have finished the proof of the claim, and the proof of the lemma by taking $\phi = \varphi$.
\end{proof}

\boolMethodTwo*

\begin{proof}
($\Leftarrow$) Assuming that $\varphi_e$ is valid, we must show that $\mueval{\sigma(\varphi)} = 1$ for any instantiation $\sigma$ and valuation $\valuation$. Given a pair of these, we define an extended valuation $\valuation'$ with $\valuation'(\ctxap) = \feval{\sigma(\Context{c}{\ctxarg})}$ for every $c[\psi] \in \consub{\varphi}$, and $\valuation'(p) = \valuation(p)$ otherwise. First, we have $\feval{\sigma(\phi)} = \feval[\valuation']{\decon{\phi}}$ for every subformula $\phi$ by the substitution lemma as applied in \cref{clm:bool:arg}. Using this identity, if the conclusion of the top implication in $\varphi$ held, the we would have finished the proof with $\feval{\sigma(\varphi)} = \feval[\valuation']{\decon{\varphi}} = 1$. Therefore, we must prove that the premise of the top implication in $\varphi_e$ holds. Using $\feval[\valuation']{\sigma(\ctxarg_i)} = \feval[\valuation']{\decon{\ctxarg_i}}$ and the definition of $\valuation'$, each clause of its conjunction reads
\begin{align*}
(\feval[\valuation']{\sigma(\ctxarg_1)} \to \feval[\valuation']{\sigma(\ctxarg_2)}) &\to (\feval[\valuation']{\sigma(c)[\hole/\sigma(\ctxarg_1)]} \to \feval[\valuation']{\sigma(c)[\hole/\sigma(\ctxarg_2)]}) \end{align*}
and this is true by the monotonicity of \cref{lem:bool:monotonic}.

($\Rightarrow$) Suppose $\varphi$ is valid, then $\feval{\sigma(\varphi)} = 1$ for every $\sigma$ and $\valuation$, in particular for $\cninst_\varphi$. We must prove that $\feval{\varphi_e} = 1$ holds for every valuation $\valuation$. Fixed a valuation $\valuation$, if the premise of $\varphi_e$ does not hold, then $\varphi_e$ is trivially satisfied. Otherwise, we need $\feval{\decon{\varphi}} = 1$ and since $\feval{\sigma(\varphi)} = 1$, we will prove $\feval{\decon{\phi}} = \feval{\sigma(\phi)}$ by induction on $\phi$. For subformulas without context variables, $\sigma(\phi) = \decon{\phi}$, so the equality holds trivially. Otherwise, consider the substitution $\gamma(\ctxap) = \cninst(\Context{c}{\ctxarg})$ for $\Context{c}{\ctxarg} \in \consub{\phi}$, which satisfies $\gamma(\decon{\phi}) = \sigma(\phi)$. The substitution lemma tells us that $\feval{\sigma(\phi)} = \feval{\gamma(\decon{\phi})} = \feval[\valuation']{\decon{\phi}}$ for some $\valuation'$ such that $\valuation'(\ctxap) = \feval{\cninst_\varphi(\Context{c}{\ctxarg})}$. If we show $\feval{\cninst_\varphi(\Context{c}{\ctxarg})} = \valuation(\ctxap)$, then $\valuation = \valuation'$ and we will be done. Expanding $\cninst_\varphi(\Context{c}{\psi})$,
\[ \kern-1.5ex \bigwedge_{\Context{c}{\otherarg} \in \consub{\varphi}} (\feval{\cninst_\varphi(\ctxarg)} \rightarrow \feval{\decon{\otherarg}}) \to \valuation(\ctxap{c}{\otherarg}) \stackrel{\text{(IH)}}{=} \kern-2ex \bigwedge_{\Context{c}{\otherarg} \in \consub{\varphi}} (\feval{\decon{\ctxarg}} \rightarrow \feval{\decon{\otherarg}}) \to \valuation(\ctxap{c}{\otherarg}) \]
where we have applied induction hypothesis to the subformula $\psi$ of $\Context{c}{\psi}$. Let us consider two cases for each clause depending on whether $\feval{\decon{\ctxarg}} \rightarrow \feval{\decon{\otherarg}}$ is satisfied.
\begin{itemize}
	\item If it is not, the complete implication trivially holds, so the clause can be ignored from the conjunction.
	\item Otherwise, the clause evaluates to $\valuation(\ctxap{c}{\otherarg})$. Moreover, the premise holds, and we can use it on the additional condition of $\varphi_e$ to conclude $\valuation(\ctxap) \rightarrow \valuation(\ctxap{c}{\otherarg})$.
\end{itemize}
This means every clause values $\valuation(\ctxap)$ (this happens with $\otherarg = \psi$) or $\true$, so the value of the whole conjunction is $\valuation(\ctxap)$, as we wanted to prove.
\end{proof}

\subsection{\textmu-calculus}

\subsmuvars*

\begin{proof}
By induction on the formula,
\begin{itemize}
	\item ($p$) There is nothing to prove because $\sigma(p) = p$.
	\item ($X$) $\mueval[\eta]{\sigma(X)} = \eta'(X) = \mueval[\eta']{X}$, by definition of $\eta'$ and by the semantics of $\mu$-calculus, respectively.
	\item ($\mu X. \phi$) We have
\begin{align*}
	\mueval{\sigma(\mu X. \phi)} &= \mueval{\mu X. \sigma|_{V \backslash \{X\}}(\phi)} = \bigcap \{ U \subseteq S \mid \mueval[{\eta[X/U]}]{\sigma|_{V \backslash \{X\}}(\phi)} \subseteq U \} \\
	&\stackrel{(*)}{=} \bigcap \{ U \subseteq S \mid \mueval[{\eta'[X/U]}]{\phi} \subseteq U \} = \mueval[\eta']{\mu X. \phi}
\end{align*}
where $(*)$ is proven in the following. By induction hypothesis, $\mueval[{\eta[X/U]}]{\sigma|_{V \backslash \{X\}}(\phi)} = \mueval[\eta_\phi]{\phi}$ for some $\eta_\phi$ such that $\eta_\phi(Y) = \mueval[{\eta[X/U]}]{\sigma|_{V \backslash \{X\}}(Y)}$. If $Y = X$, the result is $U$. If $Y \neq X$, $\eta_\phi(Y) = \mueval[{\eta[X/U]}]{\sigma(Y)} = \mueval[\eta]{\sigma(Y)} = \eta'(Y)$, because $\sigma(Y)$ does not depend on $X$ (it should be either $Y$ or closed). Then, $\eta_\phi = \eta_\phi[X/U] = \eta'[X /U]$ and we have finished the proof.
	\item ($\nu X. \phi$) The proof is essentially the same as that of $\mu X. \phi$.
	\item All other cases hold homomorphically.
\end{itemize}
\end{proof}

\subsmuprops*

\begin{proof}
We define $\eta \valbound{\varphi} \eta'$ if $\eta(X) \subseteq \eta'(X)$ for all $X \in V$. We will prove $\mueval[\mathcal{K}', \mint]{\varphi} = \mueval[\mathcal{K},\mint]{\sigma(\varphi)}$ by induction on the stronger statement $\mueval[\mathcal{K}', \eta]{\phi} \subseteq \mueval[\mathcal{K}, \eta]{\sigma(\phi)}$ if $\mint \valbound{\phi} \eta$ and $\mueval[\mathcal{K}, \eta]{\sigma(\phi)} \subseteq \mueval[\mathcal{K}', \eta]{\phi}$ if $\eta \valbound{\phi} \mint$ for any formula $\phi$.
\begin{itemize}
	\item ($p$) We have $\mueval[\mathcal{K}, \mint]{\sigma(p)} = \ell'(p) = \mueval[\mathcal{K}', \eta]{p}$ by hypothesis. On the other hand, by the monotonicity of $\mu$-calculus, $\mueval[\mathcal{K}, \eta]{\sigma(p)} \subseteq \mueval[\mathcal{K}, \mint]{\sigma(p)}$ if $\eta \subseteq \mint$ and vice versa, so the statement holds in either case.
	\item ($\neg\, p$) $\sigma(\neg\, p) = \neg\, p$ by the restriction in the statement, so there is nothing to prove.
	\item ($X$) $\sigma(X) = X$, so there is nothing to prove.
	\item ($\mu X. \phi$) One side is $\mu X. \phi$ and the other one is $\sigma(\mu X. \phi) = \mu X. \sigma(\phi)$. In case $\mint \valbound{\phi} \eta$, we have to prove
\begin{align*}
	\mueval[\mathcal{K}', \eta]{\mu X. \phi} &= \bigcap \{ U \subseteq S \mid \mueval[\mathcal{K}', {\eta[X/U]}]{\phi} \subseteq U \} \\
	&\subseteq \bigcap \{ U \subseteq S \mid \mueval[{\mathcal{K}, \eta[X/U]}]{\sigma(\phi)} \subseteq U \} = \mueval[\mathcal{K}, \eta]{\mu X. \sigma(\phi)}
\end{align*}
Observe that $\mint(X) = \mueval[\mathcal{K}, \mint]{\mu X. \sigma(\phi)} \subseteq \mueval[\mathcal{K}, \eta]{\mu X. \sigma(\phi)}$ by monotonicity, so every set $U$ in the second intersection must satisfy $\mint(X) \subseteq U$. Then, $\mint \subseteq \eta[X/U]$ and $\mueval[{\mathcal{K}', \eta[X/U]}]{\phi} \subseteq \mueval[\mathcal{K}, {\eta[X/U]}]{\sigma(\phi)}$. If $\mueval[{\mathcal{K}, \eta[X/U]}]{\sigma(\phi)} \subseteq U$, then $\mueval[{\mathcal{K}', \eta[X/U]}]{\phi} \subseteq U$, so every $U$ in the second intersection is also in the first one, and the inclusion holds.
For the opposite case, $\eta \valbound{\phi} \mint$, we need
\begin{align*}
	\mueval[\mathcal{K}, \eta]{\mu X. \sigma(\phi)} &= \bigcap \{ U \subseteq S \mid \mueval[{\mathcal{K}, \eta[X/U]}]{\sigma(\phi)} \subseteq U \} \\
	&\subseteq \bigcap \{ U \subseteq S \mid \mueval[\mathcal{K}', {\eta[X/U]}]{\phi} \subseteq U \} = \mueval[\mathcal{K}', \eta]{\mu X. \phi}
\end{align*}
Now, $\mueval[\mathcal{K}, \eta]{\mu X. \sigma(\phi)} \subseteq \mueval[\mathcal{K}, \mint]{\mu X. \sigma(\phi)} = \mint(X)$, so only the subsets of $\mint(X)$ are interesting for calculating the intersection (indeed, the fixpoint must be one of these sets). If $U \subseteq \mint(X)$, then $\eta[X/U] \subseteq \mint$, $\mueval[\mathcal{K}, {\eta[X/U]}]{\sigma(\phi)} \subseteq \mueval[{\mathcal{K}', \eta[X/U]}]{\phi}$, and so every such set in the second intersection is also in the first one.

	\item ($\nu X. \phi$) One side is $\nu X. \phi$ and the other one is $\sigma(\nu X. \phi) = \nu X. \sigma(\phi)$. In case $\mint \valbound{\phi} \eta$, we have to prove
\begin{align*}
	\mueval[\mathcal{K}', \eta]{\nu X. \phi} &= \bigcup \{ U \subseteq S \mid U \subseteq \mueval[{\mathcal{K}', \eta[X/U]}]{\phi} \} \\
	&\subseteq \bigcup \{ U \subseteq S \mid U \subseteq \mueval[{\mathcal{K}, \eta[X/U]}]{\sigma(\phi)} \} = \mueval[\mathcal{K}, \eta]{\nu X. \sigma(\phi)}
\end{align*}
Again, we observe that $\mint(X) = \mueval[\mathcal{K}, \mint]{\nu X. \sigma(\phi)} \subseteq \mueval[\mathcal{K}, \eta]{\nu X. \sigma(\phi)}$, so only the supersets of $\mint(X)$ are interesting for the computation of the union (the fixpoint should be one of them). Then, if $\mint(X) \subseteq U$, we have $\mint \subseteq \eta[X/U]$, so $\mueval[{\mathcal{K}', \eta[X/U]}]{\phi} \subseteq \mueval[\mathcal{K}, {\eta[X/U]}]{\sigma(\phi)}$. Every $U$ in the first union is also in the second one, so the inclusion holds. The opposite case is analogous.
	\item All other cases hold homomorphically.
\end{itemize}
\end{proof}

\mucMint*

\begin{proof}
The relation of being a subformula is a partial order, which can be arbitrarily extended to a linear order $X_1, \ldots, X_n$,	 so the fixpoint subformulas $\alpha X_k. \phi_k$ of $\varphi$ can be arranged in a sequence such that the only free variables $X_l$ in $\alpha X_k. \phi_k$ satisfy $l < k$. Let $\eta_0$ be a partial function that is not defined anywhere. For each $1 \leq k \leq n$, if $n$ is the number of fixpoint subformulas, we can extend $\eta_{k-1}$ with $\eta_k(X_k) = \mueval[\eta_{k-1}]{\alpha X_k. \phi_k}$, which is well-defined since the only free variables of $\alpha X_k. \phi_k$ are already defined in $\eta_{k-1}$. $\eta_n$ is the $\eta$ of the statement.

Moreover, every $\eta'$ satisfying the same conditions coincides with $\mint$ in $X_1, \ldots, X_n$. In effect, for $1 \leq k \leq n$, assume $\eta'(X_l) = \mint(X_l)$ for all $l < k$. Then, $\eta'(X_k) = \mueval[\eta']{\alpha X_k. \phi_k} = \mueval[\mint]{\alpha X_k. \phi_k} = \mint(X_k)$ because $\phi_k$ only contains free occurrences of $X_l$ for $l < k$ and $\eta'$ and $\mint$ coincide on them.
\end{proof}

\mucMintsubs*

\begin{proof}
The inductive construction of $\sigma$ is similar to that of \cref{lem:mucalc:mint}. For any ascending ordering of the fixpoint subformulas $\alpha X_k. \phi_k$ of $\varphi$, we define the partial function $\sigma_0$ that is not defined anywhere. Then, for each $1 \leq k \leq n$, we extend $\sigma_{k-1}$ by $\sigma_k(X_k) = \alpha X_k. \sigma_{k-1}(\phi_k)$. Finally, we set $\sigma = \sigma_n$. Moreover,
\begin{itemize}
	\item $\sigma_k(X_k)$ is closed, since $\phi_k$ only contain free variables $X_l$ for $l < k$, but they are replaced by closed formulas by $\sigma_{k-1}$;
	\item $\sigma_k(X_k) = \alpha X_k. \sigma_{k-1}(\phi_k) = \sigma_{k-1}(\alpha X_k. \phi_k) = \sigma_k(\alpha X_k. \phi_k)$, where $\sigma_{k-1}$ can be replaced by $\sigma_k$ because $X_k$ is not free in the argument;
	\item $\sigma'(X_k) = \sigma'(\alpha X_k. \phi_k) = \alpha X_k. \sigma'|_{V \setminus \{X_k\}}(\phi_k) = \alpha X_k. \sigma_{k-1}(\phi_k) = \sigma_k(X_k)$ for any other $\sigma'$ in the conditions of the statement, because $\phi_k$ only contains variables up to $k$, where $\sigma'|_{V \setminus \{X_k\}}$ and $\sigma_{k-1}$ coincide by induction hypothesis;
	\item $\mueval[\eta']{\sigma_k(\phi)} = \mueval{\phi}$ for a subformula $\phi$ that only contains variables up to $X_{k}$ can be proved by induction on the syntax tree, where most cases follow homomorphically. For a variable, $X_l$, if $l < k$, $\sigma_k(X_l) = \sigma_{k-1}(X_l)$ and we apply induction hypothesis on $k$. Otherwise, if $l = k$, $\mueval[\eta']{\sigma_k(X_k)} = \mueval[\eta']{\sigma_k(\alpha X_k. \phi_k)} = \mueval[\eta']{\sigma_{k-1}(\alpha X_k. \phi_k)} = \mueval{\alpha X_k. \phi_k}$ by induction hypothesis on $k$. The case $\mu_{X_l} X_l. \phi_l$ is trivial since $\sigma_k(\mu_{X_l} X_l. \phi_l) = \sigma_{k-1}(\mu_{X_l} X_l. \phi_l)$ for all $l \leq k$.

\end{itemize}
These properties are maintained by extensions, since no variable of higher index occurs in the formulas.
\end{proof}

\begin{lemma} \label{lem:mucalc:ag}
For every formula $\varphi \in \mucx$, $s \in \mueval{\AG \varphi}$ iff $s' \in \mueval{\varphi}$ for all $s \to^* s'$.
\end{lemma}

\begin{proof}
Remember that $\AG \varphi \equiv \nu X. [.] X \wedge \varphi$. Then, $s \in \mueval{\AG \varphi} = \mueval[{\eta[X/\mueval{\AG \varphi}]}]{[.] X} \cap \mueval{\varphi}$, where the $[X/\mueval{\AG \varphi}]$ can be dropped in the second evaluation because $X$ does not appear in $\varphi$. By the semantics of $\mu$-calculus, this means $s \in \mueval{\varphi}$ and $s' \in \mueval{\AG\varphi}$ for all $s \to s'$. It is easy to see using this, by induction on the length of $s \to^* s'$, that the statement holds.
\end{proof}

Let $\Gamma : \mathcal{P}(S) \to \mathcal{P}(S)$ be $\Gamma(U) = \{ s \in U \mid s' \in U \text{ for all } s \to^* s' \}$. \Cref{lem:mucalc:ag} can be stated as $\mueval{\AG \varphi} = \Gamma(\mueval{\varphi})$. It is easy to prove that $\Gamma(\emptyset) = \emptyset$, $\Gamma(S) = S$, $\Gamma(\Gamma(U)) = \Gamma(U)$, and that $\Gamma$ is monotonic.

\mucGamma*

\begin{proof}
We restrict without loss of generality to the states $s'$ of the Kripke structure reachable $s \to^* s'$ from $s$. Then, by \cref{lem:mucalc:ag}, $\mueval[\mint]{\psi} \subseteq \mueval[\mint]{\psi'}$. Since $\hole$ always appears unnegated in $\varphi$ we can consider it as a atomic proposition to apply the propositional substitution lemma (\cref{lem:mucalc:subsap}). The substitutions $\sigma_1(\hole) = \psi$ and $\sigma_2(\hole) = \psi'$ (extended as identities otherwise) give us the terms $\mueval[\mint]{\varphi[\hole/\psi]} = \mueval[{\mint[\hole/\mueval[\mint]{\psi}]}]{\varphi}$ and $\mueval[\mint]{\varphi[\hole/\psi']} = \mueval[{\mint[\hole/\mueval[\mint]{\psi'}]}]{\varphi}$ of the statement. The monotonicity of the $\mu$-calculus does the rest.
\end{proof}

\mucalcModelPrime*

\begin{proof}
Consider the valuation $\mint$ obtained by applying \cref{lem:mucalc:mint} on $\sigma(\varphi)$, we will prove that $\mathcal{K}' \vDash \cninst_\varphi(\varphi)$ where $\mathcal{K}' \coloneq (S, \to, I, \AP', \ell')$, $\AP' \coloneq \AP \cup \{\ctxap \mid \Context{c}{\ctxarg} \in \consub{\varphi} \}$, $\ell'(p) = \ell(p)$ if $p \in  \AP$, and $\ell'(\ctxap) = \mueval[\mint]{\sigma(\Context{\context}{\ctxarg})}$. Observe that $\mueval[\mathcal{K}']{\sigma(\varphi)} = \mueval[\mathcal{K}]{\sigma(\varphi)}$, because $\ctxap$ does not occur in $\sigma(\varphi)$, so we assume without loss of generality that $\mathcal{K} = \mathcal{K}'$ and drop the Kripke structure subindex for the rest of the proof.

\begin{claim}
$\mueval[\mint]{\sigma(\phi)} = \mueval[\mint]{\decon{\phi}} = \mueval[\mint]{\devarf{\phi}}$ for any subformula $\phi$ of $\varphi$. \label{clm:arg}
\end{claim}

\noindent $\phi$ is transformed (1) to $\decon{\phi}$ by replacing $\Context{\context}{\ctxarg}$ by $\ctxap$, and then (2) to $\devarf{\phi}$ by applying the closed substitution $\msigma$ of \cref{lem:mucalc:mintsubs} for $\sigma(\varphi)$. For (1), \cref{lem:mucalc:mintsubs} and $\devarf{\phi} = \hat\sigma(\decon{\phi})$ imply $\mueval[\mint]{\devarf{\phi}} = \mueval[\mint]{\msigma(\decon{\phi})} = \mueval[\mint]{\decon{\phi}}$ because $\mint(X) = \mueval[\mint]{\msigma(X)}$. For (2), we consider $\sigma'(\ctxap) = \sigma(\Context{c}{\ctxarg})$, which satisfies $\sigma'(\decon{\varphi}) = \sigma(\varphi)$ and $\ell'(\ctxap) = \mueval[\mint]{\sigma(\Context{c}{\ctxarg})}$ by definition of $\ell'$. After that, \cref{lem:mucalc:subsap} let us conclude $\mueval[\mint]{\sigma(\varphi)} = \mueval[\mint]{\sigma'(\decon{\varphi})} = \mueval[\mint]{\decon{\varphi}}$.

\begin{claim}
$\mueval[\mint]{\sigma(\varphi)} = \mueval[\mint]{\cninst_\varphi(\varphi)}$ for any subformula $\phi$ of $\varphi$. \label{clm:final}
\end{claim}

\noindent We proceed by induction, like in \cref{thm:bool:method2}. Clearly, for the subformulas without context variables, $\sigma(\phi) = \phi = \cninst_\varphi(\phi)$ and the property trivially holds. Otherwise, assume that $\mueval[\mint]{\sigma(\phi')} = \mueval[\mint]{\cninst_\varphi(\phi')}$ for every subformula $\phi'$ of $\phi$. First, we consider subformulas $\Context{c}{\ctxarg}$,
\begin{align*}
	\textstyle \mueval[\mint]{\cninst_\varphi(\Context{\context}{\ctxarg})} &= \textstyle \mueval[\mint]{\bigwedge_{\Context{c}{\otherarg} \in \consub{\varphi}} \AG (\cninst_\varphi(\ctxarg) \rightarrow \devarf{\otherarg}) \to \ctxap{c}{\otherarg}} \\
	&= \textstyle \bigcap_{\Context{c}{\otherarg} \in \consub{\varphi}} S \setminus \Gamma(S \setminus \mueval[\mint]{\cninst_\varphi(\ctxarg)} \cup \mueval{\decon{\otherarg}}) \cup \ell(\ctxap{c}{\otherarg}) \\
	&= \textstyle \bigcap_{\Context{c}{\otherarg} \in \consub{\varphi}} S \setminus \Gamma(S \setminus \mueval[\mint]{\cninst_\varphi(\ctxarg)} \cup \mueval[\mint]{\sigma(\otherarg)}) \cup \ell(\ctxap{c}{\otherarg}) & \text{(by \cref{clm:arg})} \\
	&= \textstyle \bigcap_{\Context{c}{\otherarg} \in \consub{\varphi}} S \setminus \Gamma(S \setminus \mueval[\mint]{\sigma(\ctxarg)} \cup \mueval[\mint]{\sigma(\otherarg)}) \cup \ell(\ctxap{c}{\otherarg}) & \text{(by IH)} \\
	&= \textstyle \bigcap_{\Context{c}{\otherarg} \in \consub{\varphi}} S \setminus \mueval[\mint]{\AG ({\sigma(\ctxarg) \to \sigma(\otherarg)})} \cup \ell(\ctxap{c}{\otherarg})
\end{align*}
Notice we only apply induction hypothesis to the subformula $\psi$ of $\Context{c}{\psi}$, but \cref{clm:arg} can be applied to any formula, even if not a subformula of $\Context{c}{\psi}$.
Our goal is to prove that $\mueval[\mint]{\cninst_\varphi(\context[\ctxarg])} = \mueval[\mint]{\sigma(\context[\ctxarg])}$ ($=\ell(\ctxap)$).
The clause for $\otherarg = \psi$ in the intersection is exactly $\ell(\ctxap) = \mueval[\mint]{\sigma(\context[\ctxarg])}$, since the $\mathbf{AG}$-formula is a tautology. Thus, $\mueval{\cninst_\varphi(\context[\ctxarg])} \subseteq \mueval[\mint]{\sigma(\context[\ctxarg])}$.
It remains to prove that this is an equality, i.e.\ that $s \in \mueval{\ctxap}$ belongs to every clause of the intersection. If it does not belong to the first argument of the union, we must show $s \in \ell(\ctxap{c}{\otherarg})$, but we have $s \in \mueval[\mint]{\AG (\psi \to \otherarg)}$. Since $s \in \ell(\ctxap) = \mueval[\mint]{\sigma(\context[\ctxarg])}$, we also have $s \in \ell(\ctxap{c}{\otherarg}) = \mueval[\mint]{\sigma(\context[\otherarg])}$ by \cref{lem:mucalc:monotonic}, and $s$ is in the intersection.

This can be extended to arbitrary formulas with propositional substitution lemma (\cref{lem:mucalc:subsap}) using $\sigma'(\ctxap) = \cninst_\varphi(\Context{c}{\ctxarg})$ as substitution. Since $\sigma'(\decon{\phi}) = \cninst_\varphi(\phi)$, we have $\mueval[\mint]{\cninst_\varphi(\phi)} = \mueval[\mint]{\sigma'(\decon{\phi})} = \mueval[\mint]{\decon{\phi}}$. The substitution lemma can be applied because $\mint(X) = \mueval[\mint]{\alpha X. \sigma(\phi_X)} = \mueval[\mint]{\alpha X. \cninst_\varphi(\phi_X)}$ for every subformula $\alpha X. \phi_X$ of $\phi$ by induction hypothesis. Moreover, $\ell(\ctxap) = \mueval[\mint]{\cninst_\varphi(\context[\ctxarg])}$, which we have just proven for the $\ctxap$ that occur in $\phi$.

The conclusion follows from \cref{clm:final} with $\phi = \varphi$, since $\varphi$ and $\sigma(\varphi)$ are closed.
\end{proof}

\mucalcModelsTwo*

\begin{proof}
(1) Given $\mathcal{K}$, $\varphi$, and $\sigma$, let $\mint$ be the valuation of \cref{lem:mucalc:mint} for $\sigma(\varphi)$. We define an extended Kripke structure $\mathcal{K'}$ with $\ell'(\ctxap) = \mueval[\mint]{\sigma(\Context{c}{\ctxarg})}$, $\ell(p_{X}) = \mint(X)$, and $\ell'(p) = \ell(p)$ otherwise. We claim that the premise of the top implication of $\varphi_e$ holds.
\[ \bigwedge_{\Context{c}{\psi_1}, \Context{c}{\psi_2} \in \consub{\varphi}} \kern-1.5ex \AG (\AG (\devarp{\ctxarg_1} \to \devarp{\ctxarg_2}) \to (\ctxap{c}{\ctxarg_1} \to \ctxap{c}{\ctxarg_2})) \wedge \bigwedge_{X \in F} \AG (p_{X} \leftrightarrow \alpha X. \phi_{X}^+) \]
First, we have $\mueval[\mint]{\devarp{\phi}} = \mueval[\mint]{\decon{\phi}}$ by \cref{lem:mucalc:subsvar} with variable substitution $\chi(X) = p_X$, since $\ell'(p_X) = \mint(X)$. We also have $\mueval[\mint]{\sigma(\phi)} = \mueval[\mint]{\decon{\phi}}$ from \cref{lem:mucalc:subsap} with substitution $\sigma'(\ctxap) = \Context{c}{\ctxarg}$ by the definitions of $\ell'$ and $\mint$. Hence, the second conjunction can be reduced to $\ell'(p_X) = \mueval[\mint]{\alpha X. \sigma(\phi_X)} = \mueval[\mint]{\sigma{(\alpha X. \phi_X)}} = \mueval[\mint]{\devarp{(\alpha X. \phi_X)}} = \mueval[\mint]{\alpha X. \devarp{\phi_X}}$ and it holds. Similarly, the first one can be reduced with previous identities and $\ell'(\ctxap) = \mueval[\mint]{\sigma(\Context{c}{\ctxarg})} = \mueval[\mint]{\sigma(c)[\hole/\ctxarg]}$, so that each clause of the inner conjunction reads
\[
\mueval[\mint]{\AG (\sigma(\ctxarg_1) \to \sigma(\ctxarg_2))} \cup (S \setminus \mueval[\mint]{\sigma(c)[\hole/\sigma(\ctxarg_1)]} \cup \mueval[\mint]{\sigma(c)[\hole/\sigma(\ctxarg_2)]}))
\]
Either a state $s$ belongs to the first set in the union or it satisfies the condition of \cref{lem:mucalc:monotonic} that implies it belongs to the second. In either case, the clause holds everywhere, so does the first conjunction, and then the whole premise. Once the premise is satisfied, the condition of $\varphi_e$ is $\decon{\varphi}$, but we know $\mueval[\mint]{\sigma(\varphi)} = \mueval[\mint]{\decon{\varphi}}$. Hence, $\mathcal{K} \models \sigma(\varphi)$ iff $\mathcal{K} \models \varphi_e$.

(2) For convenience, and without loss of generality, we can assume that all states in $\mathcal{K}$ are reachable from the initial ones. Let $\varphi_e = \varphi_p \to \decon{\varphi}$ and $\mint$ be the valuation obtained from \cref{lem:mucalc:subsvar} for $\cninst_\varphi(\varphi)$. Our goal is proving $\mueval[]{\devarp{\varphi}} = \mueval[]{\cninst_\varphi(\varphi)}$, from which the statement follows. However, this requires proving $\mint(X) = \ell'(p_X)$ and $\ell'(\ctxap) = \mueval[\mint]{\cninst_\varphi(\Context{c}{\ctxarg})}$ for applying the usual substitutions, and the premise $\varphi_p$ will help to do so.

The second part of the premise states that $p_{X} \leftrightarrow \alpha X. \devarp{\phi_{X}}$ must be satisfied in every (reachable) state. Consider the substitution $\chi(X) = p_{X}$, it follows that $\mueval[\mint]{\chi(X)} = \ell(p_{X}) = \mueval[\mint]{\alpha X. \decon{\phi_X}}$ where $\alpha X. \phi_X$ is the subformula binding $X$ in $\varphi$. Indeed, if we order the fixed point subformulas of $\varphi$ like in \cref{lem:mucalc:subsvar}, we have $\devarp{\phi_{X}} = \decon{\phi_{X}}$ for the fixpoint subformulas that do not contain free variables, and then $\ell(p_{X}) = \mueval[\mint]{\alpha X. \decon{\phi_{X}}}$ by the last conjunct. For any other variable $X$, we have $\ell(p_{X'}) = \mint(X')$ for all $X' \in \FV{\alpha X. \phi_X}$, so \cref{lem:mucalc:subsvar} let us conclude $\ell(p_{X}) = \mueval[\mint]{\alpha X. \decon{\phi_X}}$ in general.

Looking at the first conjunct of the premise, \cref{lem:mucalc:subsvar} let us obtain $\mueval[\mint]{\devarp{\ctxarg}} = \mueval[\mint]{\decon{\ctxarg}}$ with the previous $\chi$ substitution. On the other hand,
\[ \cninst_\varphi(\Context{c}{\ctxarg}) = \bigwedge_{\Context{c}{\otherarg} \in \consub{\varphi}} \AG (\cninst_\varphi(\ctxarg) \to \decon{\otherarg}) \to \ctxap{c}{\otherarg} \]
We proceed by induction. For subformulas without contexts, $\cninst_\varphi(\ctxarg) = \ctxarg = \decon{\ctxarg}$. Then, $\ell(\ctxap) \subseteq \mueval[\mint]{\AG (\cninst_\varphi(\ctxarg) \to \decon{\otherarg}) \to \ctxap{c}{\otherarg}}$, because either $s \in \ell(\ctxap)$ does not satisfy the premise and trivially belongs to the implication, or it satisfies the premise and $s \in \ell(\ctxap{c}{\otherarg})$ since $\ell(\ctxap) \subseteq \ell(\ctxap{c}{\otherarg})$ because of the implication in the second disjunct. Moreover, for $\otherarg = \ctxarg$, the clause values exactly $\ell(\ctxap)$, so $\mueval{\cninst_\varphi(\Context{c}{\ctxarg})} = \ell(\ctxap)$. This identity can be inductively extended to all pairs $(k, l)$ following the order of formulas and using that $\mueval[\mint]{\cninst_\varphi(\ctxarg)} = \mueval[\mint]{\decon{\ctxarg}}$ as follows from \cref{lem:mucalc:subsap}, whose premises hold by induction hypothesis.

Finally, we can then apply \cref{lem:mucalc:subsap} to conclude that $\mueval[\mint]{\cninst_\varphi(\varphi)} = \mueval[\mint]{\decon{\varphi}}$, which implies $\mathcal{K}' \models \varphi_e$ iff $\mathcal{K}' \models \cninst_\varphi(\varphi)$.
\end{proof}

\subsection{CTL and LTL}

\ctlMain*

\begin{proof}A straightforward induction shows that $\ctltomc{(\sigma(\varphi))} = \ctltomc{\sigma}(\ctltomc{\varphi})$ for any instantiation $\sigma$. Moreover, observe that $\ctltomc{\cninst_\varphi}$ for the canonical context $\cninst_{\varphi}$ of this \lcnamecref{thm:ctl:main} coincides with the canonical context $\cninst_{\ctltomc{\varphi}}$ of \cref{thm:mucalc:main}. Similarly, $\ctltomc{\varphi_e}$ for the $\varphi_e$ of this \lcnamecref{thm:ctl:main} coincides with the $(\ctltomc{\varphi})_e$ of \cref{thm:mucalc:method2}, since the translation does not introduce any free $\mu$-calculus variable in fixpoint formulas and the last conjunction disappears.

(1) $\Leftrightarrow$ (2). If $\varphi$ is valid, $\cninst_\varphi(\varphi)$ is valid by \cref{def:satisfaction}, so we will focus on the other direction. Suppose $\cninst_\varphi(\varphi)$ is a valid CTL formula,
then $\ctltomc{(\cninst_\varphi(\varphi))}$ is a valid $\mu$-calculus formula.

\begin{center}
\begin{tikzpicture}[>=latex, x=6em, y=3.5em]
	\node (CC) at (0, 1) {$\varphi$};
	\node (C) at (1, 1) {$\sigma_c(\varphi)$};
	\node (CM) at (0, 0) {$\ctltomc{\varphi}$};
	\node (M) at (1, 0) {$\ctltomc{\sigma_c}(\ctltomc{\varphi})$};
	\node (XC) at (-1, 1) {$\cninst_\varphi(\varphi)$};
	\node (XM) at (-1, 0) {$\cninst_{\ctltomc{\varphi}}(\ctltomc{\varphi})$};

	\node (SC) at (2, 1) {$\sigma$};
	\node (SM) at (2, 0) {$\ctltomc{\sigma}$};

	\node at (-2, 1) {CTL};
	\node at (-2, 0) {$\mu$-calculus};

	\draw[->] (CM) -- (M);
	\draw[->] (CC) -- (C);
	\draw[->] (CM) -- (XM);
	\draw[->] (CC) -- (XC);
	\draw[->] (CC) -- (CM);
	\draw[->] (C) -- (M);
	\draw[->] (SC) -- (SM);
	\draw[->] (XC) -- (XM);
\end{tikzpicture}
\end{center}

\noindent We have $\ctltomc{(\cninst_\varphi(\varphi))} = \cninst_{\ctltomc{\varphi}}(\ctltomc{\varphi})$. Then, the $\mu$-calculus formula $\ctltomc{\varphi}$ is valid by \cref{thm:mucalc:main}, which implies $\ctltomc{\sigma_c}(\ctltomc{\varphi})$ is also valid. Hence, the CTL formula $\sigma_c(\varphi)$ is valid in the standard CTL semantics, since it translates to $\ctltomc{(\sigma_c(\varphi))} = \ctltomc{\sigma_c}(\ctltomc{\varphi})$.

(1) $\Leftrightarrow$ (3). As mentioned before, notice that $\varphi_e$ is the translation to CTL of the $\mu$-calculus formula $(\ctltomc{\varphi})_e$ in \cref{thm:mucalc:method2}. The translation from CTL to $\mu$-calculus does not introduce any free variable in a fixpoint formula, so the second conjunction of the premise in $(\ctltomc{\varphi})_e$ is empty. Since (1) $\Leftrightarrow$ (3) holds in the $\mu$-calculus side, it also holds in the CTL one.
\end{proof}

\lemLasso*

\begin{proof}
An LTL property is valid if it is satisfied by every path of every (finite) Kripke structure. Each such path gives a lasso, so it is enough to check properties on those models. In any such model, the semantics of LTL and $\mu$-calculus coincide. Indeed, if $\pi$ is the path through the lasso, $\mcEX \varphi = \{ s \in S \mid \exists s' \quad s \to s' \wedge s' \in \mueval{\varphi} \}$ translates to $\pi_k \in \mcEX \varphi$ iff $\pi_{k+1} \in \mueval{\varphi}$ for every $k \in \N$ ($s'$ always exists and it is unique, $\pi_{k+1}$), which coincides with the meaning of $\X \varphi$. The same happens with $\mcAX$. \end{proof}

\ltlMain*

\begin{proof}
We recall from \cref{thm:ctl:main} that $\ctltomc{\sigma(\varphi)} = \ctltomc{\sigma}(\ctltomc{\varphi})$, $\ctltomc{\cninst_\varphi} = \cninst_{\ctltomc{\varphi}}$, and $\ctltomc{\varphi_e} = (\ctltomc{\varphi})_e$.

A general schema of the proof is shown in \cref{fig:ltl:main}. $(1) \Rightarrow (2)$ trivially holds by definition. For $(2) \Rightarrow (1)$, in order for $\varphi$ to be valid, $\sigma(\varphi)$ must hold for every instantiation $\sigma$ and every lasso $\mathcal{L}$ by \cref{lem:lasso}. Given a pair of these, by the same lemma, $\mathcal{L} \models_{\text{LTL}} \sigma(\varphi)$ if $\mathcal{L}$ is a model of $\ctltomc{(\sigma(\varphi))} = \ctltomc{\sigma}(\ctltomc{\varphi})$ for the $\mu$-calculus. Moreover, \Cref{lem:mucalc:modelprime} claims that this happens if $\ctltomc{(\cninst_\varphi(\varphi))} = \cninst_{\ctltomc{\varphi}}(\ctltomc{\varphi})$ holds in a structure $\mathcal{L}'$ extended with some additional atomic propositions that maintains the shape of $\mathcal{L}$. Finally, this is reduced $\mathcal{L} \models_{\text{LTL}} \cninst_\varphi(\varphi)$ by \cref{lem:lasso}, which is true because $\varphi$ is valid in LTL.
For $(3) \Rightarrow (1)$, we follow the same argument, but then use \cref{thm:mucalc:models2} to reduce the problem to $\ctltomc{\varphi_e}$ holding in a lasso $\mathcal{L}'$ with the same shape as $\mathcal{L}$. Then, we can use through \cref{lem:lasso} that $\mathcal{L} \models_{\text{LTL}} \varphi_e$ holds because $\varphi_e$ is valid. Finally, for $(2) \Rightarrow (3)$, $\varphi_e$ is valid in LTL if $\mathcal{L}' \models_{\text{LTL}} \varphi_e$ for every lasso $\mathcal{L}'$.
If the premise of $\varphi_e$, say $\varphi_p$, does not hold, this is trivially true. Otherwise, the premise of $\ctltomc{\varphi_e} = \ctltomc{(\varphi_p \to \decon{\varphi})} = \ctltomc{\varphi_p} \to \ctltomc{(\decon{\varphi})}$ holds on $\mathcal{L}'$, so $\mathcal{L}' \models (\ctltomc{\varphi})_e$ if $\mathcal{L}' \models \cninst_{\ctltomc{\varphi}}(\ctltomc{\varphi})$ by \cref{thm:mucalc:models2}, and so if $\mathcal{L}' \models_{\text{LTL}} \cninst_\varphi(\varphi)$, which holds by hypothesis.
\end{proof}

\section{Extension to non-monotonic contexts} \label{sec:extensions}

While we have limited our exposition to formulas in negation normal form, and hence to monotonic contexts, the results of \cref{sec:boolean,sec:ctl,sec:ltl} can be generalized to the unrestricted syntax of the corresponding logics and to arbitrary contexts. It is enough to change $\to$ by $\leftrightarrow$ in a few places. For example, for the Boolean case, the canonical instantiation for propositional logic and the equisatisfiable formula would be
\[
	\cninst_\varphi(c) \coloneq \bigwedge_{l=1}^{m_k} (\hole \leftrightarrow \decon{\ctxarg}) \to \ctxap
\qquad
	\left(\bigwedge_{\substack{\Context{c}{\ctxarg_1}, \Context{c}{\ctxarg_2} \\ \in \consub{\varphi}}} (\decon{\ctxarg_1} \leftrightarrow \decon{\ctxarg_2}) \to (\ctxap{c}{\ctxarg_1} \leftrightarrow \ctxap{c}{\ctxarg_1})\right) \to \decon{\varphi}
\]
Indeed, the proofs for the Boolean case would be very similar and simpler than those in \cref{sec:boolean}. The case of $\mu$-calculus would be more subtle, since contexts like $\Context{c}{X}$ would make us introduce invalid formulas like $\hole \leftrightarrow X$ where $X$ appears both under an even and an odd number of negations.

\section{List of mutated formulas from \cref{fig:rewrite}} \label{sec:mutated}

\newcommand\holdmark{\colorbox{green}{\checkmark}}
\newcommand\doesnotholdmark{\colorbox{red}{$\times$}}

\begin{enumerate}
	\item $\Context{c}{\GF p} \equiv ((\GF p) \wedge \Context{c}{\true}) \vee \Context{c}{\false}$
	\begin{enumerate}
		\item \doesnotholdmark \enspace $\Context{c}{\GF p} \equiv ((\FG p) \wedge \Context{c}{\true}) \vee \Context{c}{\false}$
		\item \doesnotholdmark \enspace $\Context{c}{\GF p} \equiv ((\GF p) \wedge \Context{c}{\true}) \vee \Context{c}{p}$
		\item \doesnotholdmark \enspace $\Context{c}{\GF p} \equiv ((\GF p) \vee \Context{c}{\true}) \wedge \Context{c}{\false}$
		\item \doesnotholdmark \enspace $\Context{c}{\GF p} \equiv ((\GF p) \wedge \Context{c}{\false}) \vee \Context{c}{\true}$
	\end{enumerate}
	\item $\Context{c}{\FG p} \equiv ((\FG p) \wedge \Context{c}{\true}) \vee \Context{c}{\false}$
	\begin{enumerate}
		\item \doesnotholdmark \enspace $\Context{c}{\FG p} \equiv ((\GF p) \wedge \Context{c}{\true}) \vee \Context{c}{\false}$
		\item \doesnotholdmark \enspace $\Context{c}{\FG p} \equiv ((\FG p) \wedge \Context{c}{\true}) \vee \Context{c}{p}$
		\item \doesnotholdmark \enspace $\Context{c}{\FG p} \equiv ((\FG p) \vee \Context{c}{\true}) \wedge \Context{c}{\false}$
		\item \doesnotholdmark \enspace $\Context{c}{\FG p} \equiv ((\FG p) \wedge \Context{c}{\false}) \vee \Context{c}{\true}$
	\end{enumerate}
	\item $\GF \Context{c}{a \W b} \equiv (\GF \Context{c}{a \U b}) \vee ((\FG a) \wedge (\GF \Context{c}{\true}))$
	\begin{enumerate}
		\item \holdmark \enspace $\GF \Context{c}{a \W b} \equiv (\GF \Context{c}{a \W b}) \vee ((\FG a) \wedge (\GF \Context{c}{\true}))$
		\item \doesnotholdmark \enspace $\GF \Context{c}{a \W b} \equiv (\GF \Context{c}{a \U b}) \vee ((\GF a) \wedge (\GF \Context{c}{\true}))$
		\item \doesnotholdmark \enspace $\GF \Context{c}{a \W b} \equiv (\GF \Context{c}{a \U b}) \vee ((\FG a) \wedge (\Context{c}{\true}))$
		\item \doesnotholdmark \enspace $\GF \Context{c}{a \W b} \equiv (\GF \Context{c}{a \U b}) \wedge ((\FG a) \vee (\GF \Context{c}{\true}))$
	\end{enumerate}
	\item $\FG \Context{c}{a \U b} \equiv ((\GF b) \wedge \FG \Context{c}{a \W b}) \vee \FG \Context{c}{\false}$
	\begin{enumerate}
		\item \doesnotholdmark \enspace $\FG \Context{c}{a \U b} \equiv ((\FG b) \wedge \FG \Context{c}{a \W b}) \vee \FG \Context{c}{\false}$
		\item \doesnotholdmark \enspace $\FG \Context{c}{a \U b} \equiv ((\GF b) \wedge \GF \Context{c}{a \W b}) \vee \FG \Context{c}{\false}$
		\item \holdmark \enspace $\FG \Context{c}{a \U b} \equiv ((\GF b) \wedge \FG \Context{c}{a \U b}) \vee \FG \Context{c}{\false}$
		\item \doesnotholdmark \enspace $\FG \Context{c}{a \U b} \equiv ((\GF b) \wedge \FG \Context{c}{a \W b}) \vee \Context{c}{\false}$
		\item \doesnotholdmark \enspace $\FG \Context{c}{a \U b} \equiv ((\GF b) \vee \FG \Context{c}{a \W b}) \wedge \FG \Context{c}{\false}$
	\end{enumerate}
	\item $h \W \Context{c}{a \U b} \equiv (h \U \Context{c}{a \U b}) \vee \G h$
	\begin{enumerate}
		\item \doesnotholdmark \enspace $h \W \Context{c}{a \U b} \equiv (h \U \Context{c}{a \W b}) \vee \G h$
		\item \holdmark \enspace $h \W \Context{c}{a \U b} \equiv (h \W \Context{c}{a \U b}) \vee \G h$
		\item \doesnotholdmark \enspace $h \W \Context{c}{a \U b} \equiv (h \U \Context{c}{a \U b}) \vee \F h$
		\item \doesnotholdmark \enspace $h \W \Context{c}{a \U b} \equiv (h \U \Context{c}{a \U b}) \wedge \G h$
		\item \doesnotholdmark \enspace $h \W \Context{c}{a \U b} \equiv (h \U \Context{c}{a \U h}) \vee \G b$
	\end{enumerate}
	\item $\Context{c}{a \U b} \W g \equiv ((\GF b) \wedge (\Context{c}{a \W b} \W g)) \vee (\Context{c}{a \U b} \U (g \vee \G \Context{c}{\false}))$
	\begin{enumerate}
		\item \doesnotholdmark \enspace $\Context{c}{a \U b} \W g \equiv ((\FG b) \wedge (\Context{c}{a \W b} \W g)) \vee (\Context{c}{a \U b} \U (g \vee \G \Context{c}{\false}))$
		\item \holdmark \enspace $\Context{c}{a \U b} \W g \equiv ((\GF b) \wedge (\Context{c}{a \U b} \W g)) \vee (\Context{c}{a \U b} \U (g \vee \G \Context{c}{\false}))$
		\item \doesnotholdmark \enspace $\Context{c}{a \U b} \W g \equiv ((\GF b) \wedge (\Context{c}{a \W b} \U g)) \vee (\Context{c}{a \U b} \U (g \vee \G \Context{c}{\false}))$
		\item \doesnotholdmark \enspace $\Context{c}{a \U b} \W g \equiv ((\GF b) \wedge (\Context{c}{a \W b} \W g)) \vee (\Context{c}{a \W b} \U (g \vee \G \Context{c}{\false}))$
		\item \doesnotholdmark \enspace $\Context{c}{a \U b} \W g \equiv ((\GF b) \wedge (\Context{c}{a \W b} \W g)) \vee (\Context{c}{a \U b} \U (g \vee \Context{c}{\false}))$
		\item \doesnotholdmark \enspace $\Context{c}{a \U b} \W g \equiv ((\GF b) \wedge (\Context{c}{a \W b} \W g)) \vee (\Context{c}{a \U b} \U (g \vee \G \Context{c}{\true}))$
		\item \holdmark \enspace $\Context{c}{a \U b} \W g \equiv ((\GF b) \wedge (\Context{c}{a \W b} \W g)) \vee (\Context{c}{a \U b} \W (g \vee \G \Context{c}{\false}))$
	\end{enumerate}
\end{enumerate}
 \fi


\begin{thebibliography}{10}

\bibitem{BiereHMW21}
Armin Biere, Marijn Heule, Hans van Maaren, and Toby Walsh, editors.
\newblock {\em Handbook of Satisfiability - Second Edition}, volume 336 of {\em
  Frontiers in Artificial Intelligence and Applications}.
\newblock {IOS} Press, 2021.
\newblock \href {https://doi.org/10.3233/FAIA336} {\path{doi:10.3233/FAIA336}}.

\bibitem{handbookMucalc}
Julian~C. Bradfield and Igor Walukiewicz.
\newblock The mu-calculus and model checking.
\newblock In Edmund~M. Clarke, Thomas~A. Henzinger, Helmut Veith, and Roderick
  Bloem, editors, {\em Handbook of Model Checking}, pages 871--919. Springer,
  2018.
\newblock \href {https://doi.org/10.1007/978-3-319-10575-8_26}
  {\path{doi:10.1007/978-3-319-10575-8_26}}.

\bibitem{ClarkeHV18}
Edmund~M. Clarke, Thomas~A. Henzinger, and Helmut Veith.
\newblock Introduction to model checking.
\newblock In Edmund~M. Clarke, Thomas~A. Henzinger, Helmut Veith, and Roderick
  Bloem, editors, {\em Handbook of Model Checking}, pages 1--26. Springer,
  2018.
\newblock \href {https://doi.org/10.1007/978-3-319-10575-8_1}
  {\path{doi:10.1007/978-3-319-10575-8_1}}.

\bibitem{spot}
Alexandre Duret{-}Lutz, Etienne Renault, Maximilien Colange, Florian Renkin,
  Alexandre~Gbaguidi Aisse, Philipp Schlehuber{-}Caissier, Thomas Medioni,
  Antoine Martin, J{\'{e}}r{\^{o}}me Dubois, Cl{\'{e}}ment Gillard, and Henrich
  Lauko.
\newblock From {Spot} 2.0 to {Spot} 2.10: What's new?
\newblock In Sharon Shoham and Yakir Vizel, editors, {\em Computer Aided
  Verification - 34th International Conference, {CAV} 2022, Haifa, Israel,
  August 7-10, 2022, Proceedings, Part {II}}, volume 13372 of {\em Lecture
  Notes in Computer Science}, pages 174--187. Springer, 2022.
\newblock \href {https://doi.org/10.1007/978-3-031-13188-2_9}
  {\path{doi:10.1007/978-3-031-13188-2_9}}.

\bibitem{minisat}
Niklas E{\'{e}}n and Niklas S{\"{o}}rensson.
\newblock An extensible {SAT}-solver.
\newblock In Enrico Giunchiglia and Armando Tacchella, editors, {\em Theory and
  Applications of Satisfiability Testing, 6th International Conference, {SAT}
  2003. Santa Margherita Ligure, Italy, May 5-8, 2003 Selected Revised Papers},
  volume 2919 of {\em Lecture Notes in Computer Science}, pages 502--518.
  Springer, 2003.
\newblock \href {https://doi.org/10.1007/978-3-540-24605-3_37}
  {\path{doi:10.1007/978-3-540-24605-3_37}}.

\bibitem{ctltrans}
E.~Allen Emerson.
\newblock Temporal and modal logic.
\newblock In Jan van Leeuwen, editor, {\em Handbook of Theoretical Computer
  Science, Volume {B:} Formal Models and Semantics}, pages 995--1072. Elsevier
  and {MIT} Press, 1990.
\newblock \href {https://doi.org/10.1016/B978-0-444-88074-1.50021-4}
  {\path{doi:10.1016/B978-0-444-88074-1.50021-4}}.

\bibitem{ctlSatisfiability}
E.~Allen Emerson and Joseph~Y. Halpern.
\newblock Decision procedures and expressiveness in the temporal logic of
  branching time.
\newblock {\em J. Comput. Syst. Sci.}, 30(1):1--24, 1985.
\newblock \href {https://doi.org/10.1016/0022-0000(85)90001-7}
  {\path{doi:10.1016/0022-0000(85)90001-7}}.

\bibitem{lipicsVersion}
Javier Esparza and Rubén Rubio.
\newblock Validity of contextual formulas.
\newblock In Rupak Majumdar and Alexandra Silva, editors, {\em 35th
  International Conference on Concurrency Theory, {CONCUR} 2024, September
  9-13, 2024, Calgary, Canada}, volume 331 of {\em LIPIcs}, pages 11:1--11:17.
  Schloss Dagstuhl - Leibniz-Zentrum f{\"{u}}r Informatik, 2024.
\newblock \href {https://doi.org/10.4230/LIPICS.CONCUR.2024.11}
  {\path{doi:10.4230/LIPICS.CONCUR.2024.11}}.

\bibitem{jacm}
Javier Esparza, Rubén Rubio, and Salomon Sickert.
\newblock Efficient normalization of linear temporal logic.
\newblock {\em J. ACM}, 71:16:1--16:42, 2024.
\newblock \href {https://doi.org/10.1145/3651152} {\path{doi:10.1145/3651152}}.

\bibitem{pysat}
Alexey Ignatiev, Ant{\'{o}}nio Morgado, and Jo{\~{a}}o Marques{-}Silva.
\newblock {PySAT}: {A} {Python} toolkit for prototyping with {SAT} oracles.
\newblock In Olaf Beyersdorff and Christoph~M. Wintersteiger, editors, {\em
  Theory and Applications of Satisfiability Testing - {SAT} 2018 - 21st
  International Conference, {SAT} 2018, Held as Part of the Federated Logic
  Conference, FloC 2018, Oxford, UK, July 9-12, 2018, Proceedings}, volume
  10929 of {\em Lecture Notes in Computer Science}, pages 428--437. Springer,
  2018.
\newblock \href {https://doi.org/10.1007/978-3-319-94144-8_26}
  {\path{doi:10.1007/978-3-319-94144-8_26}}.

\bibitem{ctlsat}
Nicola Prezza.
\newblock {CTL} ({C}omputation {T}ree {L}ogic) {SAT} solver, 2014.
\newblock URL: \url{https://github.com/nicolaprezza/CTLSAT}.

\end{thebibliography}
\end{document}